\documentstyle[11pt,epsfig]{article}
 
\def\eqn#1{eq.~(\ref{#1})}
\def\eqns#1{eqs.~#1}
\def\fig#1{fig.~{\ref{#1}}}

\def\sec#1{section~{\ref{#1}}}
\def\Sec#1{Section~{\ref{#1}}}
\def\app#1{appendix~\ref{#1}}

\def\tr{{\rm tr}}
\def\bra#1{\langle #1 |}
\def\ket#1{| #1 \rangle}

\def\braket#1#2{\langle #1 |\  #2 \rangle}
\newbox\charbox
\newbox\slabox
\def\s#1{{      
        \setbox\charbox=\hbox{$#1$}
        \setbox\slabox=\hbox{$/$}
        \dimen\charbox=\ht\slabox
        \advance\dimen\charbox by -\dp\slabox
        \advance\dimen\charbox by -\ht\charbox
        \advance\dimen\charbox by \dp\charbox
        \divide\dimen\charbox by 2
        \raise-\dimen\charbox\hbox to \wd\charbox{\hss/\hss}
        \llap{$#1$}
}}

\def\F{{\cal F}}
\def\D{{\cal D}}

\def\J{J}
\def\K{K}
\def\half{{1\over 2}}
\def\pol{\varepsilon}
\def\eps{\epsilon}

\def\Ord{{\cal O}}
\def\c{\,\cdot\,}
\def\ksl{\s k}
\def\lsl{\s \ell}

\def\si{\sigma}
\def\fermion{{\rm fermion}}
\def\scalar{{\rm scalar}}
\def\spa#1.#2{\left\langle#1\,#2\right\rangle}
\def\spb#1.#2{\left[#1\,#2\right]}
\def\lor#1.#2{\left(#1\,#2\right)}
\def\sand#1.#2.#3{%
  \left\langle\smash{#1}{\vphantom1}\right|{#2}%
  \left|\smash{#3}{\vphantom1}\right\rangle}
\def\sandp#1.#2.#3{%
  \left\langle\smash{#1}{\vphantom1}^{-}\right|{#2}%
  \left|\smash{#3}{\vphantom1}^{+}\right\rangle}
\def\sandpp#1.#2.#3{%
  \left\langle\smash{#1}{\vphantom1}^{+}\right|{#2}%
  \left|\smash{#3}{\vphantom1}^{+}\right\rangle}
\def\sandmm#1.#2.#3{%
  \left\langle\smash{#1}{\vphantom1}^{-}\right|{#2}%
  \left|\smash{#3}{\vphantom1}^{-}\right\rangle}
\def\sandpm#1.#2.#3{%
  \left\langle\smash{#1}{\vphantom1}^{+}\right|{#2}%
  \left|\smash{#3}{\vphantom1}^{-}\right\rangle}
\def\sandmp#1.#2.#3{%
  \left\langle\smash{#1}{\vphantom1}^{-}\right|{#2}%
  \left|\smash{#3}{\vphantom1}^{+}\right\rangle}
\catcode`@=11  
\newskip\humongous \humongous=0pt plus 1000pt minus 1000pt
\def\caja{\mathsurround=0pt}
\def\eqalign#1{\,\vcenter{\openup1\jot \caja
        \ialign{\strut \hfil$\displaystyle{##}$&$
        \displaystyle{{}##}$\hfil\crcr#1\crcr}}\,}
\newif\ifdtup

\newcounter{eqnumber}[section]
\renewcommand{\theeqnumber}{\thesection.\arabic{eqnumber}}
\def\equn{
\refstepcounter{eqnumber}
\eqno({\rm \theeqnumber})
}

\begin{document}
\begin{titlepage}

\begin{flushright}
UCLA/95/TEP/37\\
hep-ph/9511336 \\
November 15, 1995\\
\end{flushright}

\vskip 2.cm
\begin{center}
{\Large\bf Massive Loop Amplitudes from Unitarity}
\vskip 2cm
{\large Z. Bern\footnote{
Email address: {\tt bern@physics.ucla.edu}
} and A.G. Morgan\footnote{
Email address: {\tt morgan@physics.ucla.edu}
}
}
\vskip 0.5cm
{\it Department of Physics, University of California at Los Angeles, \\
Los Angeles, CA 90095-1547 }
\vskip 3cm                                                                    
\end{center}
\begin{abstract}
We show, for previously uncalculated examples containing a uniform
mass in the loop, that it is possible to obtain complete massive
one-loop gauge theory amplitudes solely from unitarity and known
ultraviolet or infrared mass singularities.  In particular, we
calculate four-gluon scattering via massive quark loops in QCD.  The
contribution of a heavy quark to five-gluon scattering with identical
helicities is also presented.
\end{abstract}
\vfill
\end{titlepage}

\section{Introduction}

The computation of loop amplitudes by traditional methods is
notoriously difficult involving laborious calculation.  At one loop,
traditional Feynman diagram techniques, augmented by computers, are in
wide-spread use for up to four external particles.  Beyond this,
computations become significantly more complicated and traditional
approaches are problematic.  Non-traditional approaches have recently
been used to compute all one-loop massless five-parton amplitudes
\cite{FiveGluon,Kunsztqqqqg,Fermion} and a variety of infinite
sequences of one-loop amplitudes
\cite{AllPlus,Mahlon,SusyFour,SusyOne}.  The technical developments
which allowed for this progress include methods involving spinor
helicity \cite{SpinorHelicity}, color decompositions \cite{TreeColor},
supersymmetry \cite{Susy,SusyDecomp,KunsztFourPoint}, string-theory
\cite{Long,StringBased}, recursion relations \cite{Recursive,Mahlon},
factorization \cite{TreeCollinear,AllPlus,Factorization} and unitarity
\cite{Cutting,SusyFour,SusyOne}.

These developments have been mostly limited to massless amplitudes.
However, in many physical processes, such as near particle thresholds
or in the production of heavy states, one cannot neglect masses.
There is, for example, some concern that the underlying cause of the
apparent discrepancy between the low and high energy measurements of
the strong coupling constant is due to ignoring masses \cite{Bethke}.

In this paper we extend unitarity techniques to obtain particular
examples of massive one loop amplitudes directly from tree amplitudes.
Unitarity has been a useful tool in quantum field theory since its
inception.  The Cutkosky rules \cite{Cutting,PeskinSchroeder} allow
one to obtain the imaginary (absorptive) parts of one-loop amplitudes
directly from tree amplitudes. Traditionally, dispersion relations are
then used to reconstruct real (dispersive) parts, up to ambiguities in
rational functions.  Although computationally more powerful than
Feynman rules, the rational function ambiguity has hampered the use of
unitarity as a method for obtaining complete amplitudes.

Recently, it was shown that for certain `cut-constructible' massless
amplitudes -- such as supersymmetric amplitudes -- one may obtain the
full amplitude solely from the cuts \cite{SusyFour,SusyOne}.  This
type of procedure has the inherent advantage of building loop
amplitudes from tree amplitudes which are generally compact, rather
than from more complicated Feynman diagrams.  With this technique,
massless one-loop supersymmetric amplitudes with an arbitrary number
of external legs, but fixed helicity structure, have been obtained.
The main new ingredient used to remove ambiguities from cut
constructions is the complete knowledge of all integral functions
\cite{Integrals,VNV,IntegralRecursion} entering into massless one-loop
amplitudes.  This severely limits the analytic form of amplitudes. The
cut construction method may be extended to full massless QCD by
computing to higher order in the dimensional regularization
parameter \cite{TwoLoopUnitarity,Unpublished}. In this way, all terms
in {\it any} massless one-loop amplitude become associated with branch
cuts which may be used to reconstruct the complete amplitude.

In general, one is also interested in massive theories such as QCD
with heavy quarks. The amplitudes of such theories contain logarithms
that depend only on masses; such functions do not have cuts in any
kinematic variable.  This might seem to imply that one cannot obtain
all of the terms in massive amplitudes via unitarity.  However, we
will explicitly show for four-gluon amplitudes with a massive quark
loop that the well known form of their ultraviolet or infrared-mass
singularities provides sufficient information on the coefficients of
all potentially ambiguous functions.  We also present the result for
the scattering, via a massive quark loop, of five-gluons with
identical helicities.  (These amplitudes have not been computed
previously.)  The efficiency of the method for these examples may be
characterized by the fact that intermediate algebraic expressions are
about the same size as final expressions for the amplitude.  Due to
this feature there is no need for computer assistance in performing
the computations presented in this paper.  This may be contrasted with
Feynman diagram calculations which frequently suffer from an
exponential growth in the algebra of intermediate expressions.  For
other five- and higher-point amplitudes, computer assistance is 
useful since the number of cuts and terms in the final amplitudes
proliferates.

In \sec{ReviewSection}, we briefly review a few of the established
techniques for computing amplitudes. \Sec{NaiveSection} illustrates
the ambiguities that are traditionally associated with unitarity based
calculations.  We then remove these ambiguities in
\sec{FixAmbiguitySection}.  In sections \ref{FermionLoopAmplitudes}
and \ref{RemainingSection}, we use the results of the previous
sections to calculate all four-gluon helicity amplitudes with a heavy
fermion loop. We also present the five-gluon amplitude with all
identical helicities.  Our conclusions are presented in 
\sec{SummarySection}.   We collect various useful results in 
the appendices.

\section{Review of previous techniques}
\label{ReviewSection}

We now briefly review some of the techniques which we shall use in
this paper.  Two techniques that we utilize are spinor helicity and
color decompositions.  The reader may consult the review article of
Mangano and Parke \cite{ManganoReview} for further details.

In explicit calculations with external gluons, it is usually convenient
to use a spinor helicity basis \cite{SpinorHelicity} which rewrites 
all polarization vectors in terms of massless Weyl spinors $\vert
k^{\pm} \rangle$.  In the formulation of Xu, Zhang and Chang the
plus and minus helicity polarization vectors are expressed as
$$
\pol^{+}_\mu (k;q) =  {\sandmm{q}.{\gamma_\mu}.k
      \over \sqrt2 \spa{q}.k}\, ,\hskip 1cm
\pol^{-}_\mu (k;q) =  {\sandpp{q}.{\gamma_\mu}.k
      \over \sqrt{2} \spb{k}.q} \, ,
\equn
\label{PolarizationVector}
$$  
where $q$ is an arbitrary null `reference momentum' which drops out of
the final gauge-invariant amplitudes.  
The cuts of amplitudes are also gauge invariant, so one may 
evaluate different cuts using independent choices of reference momenta.
We use the convenient notation
$$
\langle k_i^{-} \vert k_j^{+} \rangle \equiv \langle ij \rangle \, , 
\hskip 2 cm 
\langle k_i^{+} \vert k_j^{-} \rangle \equiv [ij] \, .
\equn
$$ 
These spinor products are anti-symmetric and satisfy
$$
\spa{i}.j \spb{j}.i = 2 k_i \cdot k_j  \, .
\equn
$$
This provides for an extremely compact representation of gluon amplitudes.
A useful identity is
$$
\sand{a}.{\s{\pol}^\pm(k;q)}.{b} =
\pm \frac{\sqrt{2}}{\braket{q^\mp}{k^\pm}}
\bra{a}  \left[
    \ket{q^\pm} \bra{k^\pm} + \ket{k^\mp} \bra{q^\mp}
  \right] \ket{b} \, ,
\equn\label{PolSlash}
$$
where either $\bra{a}$ or $\ket{b}$ are spinors with four-dimensional 
momenta.

For loop amplitudes, to maximize the benefit of the spinor helicity
method we must choose an appropriate regularization scheme.  In
conventional dimensional regularization \cite{CollinsBook}, the
polarization vectors are taken to be $(4-2\eps)$-dimensional; this is
incompatible with spinor helicity which assumes that the polarization
vectors are four dimensional.  An alternative is to modify the
regularization scheme to define the polarization vectors to be four
dimensional.  One also defines all internal and external states to be
four-dimensional and only continues loop momentum and phase-space integrals 
to $(4-2\eps)$ dimensions.  This is the
four-dimensional-helicity (FDH) scheme \cite{Long}, which has been
shown to be equivalent \cite{KunsztFourPoint} to an appropriate
helicity formulation of Siegel's dimensional-reduction scheme
\cite{Siegel} at one-loop.  Dimensional regularization rules, which
may be used in cut calculations of fermion loops, are reviewed in
\app{MuSlashEtc}.  The conversion between schemes has been given in
ref.~\cite{KunsztFourPoint}, so one may choose whichever scheme is
simplest to perform calculations in.  In this paper we work
exclusively in the FDH scheme but for fermion or scalar loops this
gives the same results as in the 't Hooft-Veltman
\cite{DimensionalRegularization,CollinsBook} scheme.

$SU(N_c)$ gauge theory amplitudes can be written in terms of
independent color-ordered partial amplitudes multiplied by an
associated color trace \cite{TreeColor,ManganoReview}.  One of
the key features of the partial amplitudes is that the external legs
have a fixed ordering.  For the one-loop four-gluon amplitude with a
fundamental representation loop the decomposition is%
$$
\eqalign{
{\cal A}_4^{\rm 1-loop} = g^4 
\mu_R^{2\eps}
\sum_{\sigma} &
\tr(T^{a_{\si(1)}}T^{a_{\si(2)}}T^{a_{\si(3)}}T^{a_{\si(4)}}) \, 
A_{4} (\si(1), \si(2), \si(3), \si(4))\, ,  \cr
}\equn
$$
where the sum over $\sigma$ includes all non-cyclic permutations of
the indices $\sigma(n)$ and the $T^a$ are fundamental representation
color matrices (normalized so that $\tr(T^a T^b) = \delta^{ab}$).  We
have explicitly extracted the coupling and a factor of $\mu_R^{2\eps}$
from $A_n$, where $\mu_R$ is the renormalization scale.  (We have
abbreviated the dependence of $A_{4}$ on the outgoing momenta $k_j$
and helicities $\lambda_j$ by writing the label $j$ alone.)  For
adjoint representation loops there is an analogous decomposition with
up to two color traces in each term.  A similar decomposition in
terms of `primitive amplitudes' can also be performed for the case
where some of the external particles are in the fundamental
representation \cite{Fermion}. (In refs.~\cite{FiveGluon,Fermion}, for
example, it was slightly more convenient to extract an overall factor
of $N_c$ from the color ordered amplitudes, $A_n$, but since we only
deal with fundamental representation loops in this paper there is no
need for this.)

An essential feature of this decomposition is that the partial
amplitudes, $A_{n}$, are described by diagrams with a fixed cyclic
ordering of external legs.  In this paper we deal exclusively with
such amplitudes.

\section{Naive application of Cutkosky rules}
\label{NaiveSection}

In this section we demonstrate how a naive application
of the Cutkosky rules can miss contributions to a complete one-loop
amplitude. Traditionally, one first computes the imaginary%
\footnote{By imaginary we mean the discontinuities across branch cuts.
In Feynman diagrams these are unfortunately the {\it real} parts.}
(absorptive) part of an amplitude by evaluating a four-dimensional
phase-space integral.  (The phase-space integrals are not ultraviolet
divergent and therefore converge in four-dimensions.)  The real
(dispersive) part is then reconstructed via a dispersion relation.  We
perform the cut calculation a bit differently; since we want the
complete amplitude, it is convenient to replace the phase-space
integral with an unrestricted loop momentum integral which has the
correct branch cuts \cite{SusyFour,SusyOne}.  This allows us to
simultaneously construct the real and imaginary parts.  The
ambiguities that we encounter are equivalent to those of the more
conventional approach.

We will compute cuts of massive amplitudes in all possible channels in
an attempt to reconstruct their functional form.  We consider the
amplitude, not in a physical kinematic configuration, but in a region
where exactly one of the momentum invariants is taken to be above
threshold, and the rest are negative (space-like).  In this way we
isolate cuts in a single momentum channel.  Furthermore, following
refs.~\cite{SusyFour,SusyOne}, we apply the Cutkosky rules at the
amplitude level, rather than at the diagram level.  This is
advantageous because amplitudes are generally much more compact than
diagrams.

\begin{figure}
\begin{center}
~\epsfig{file=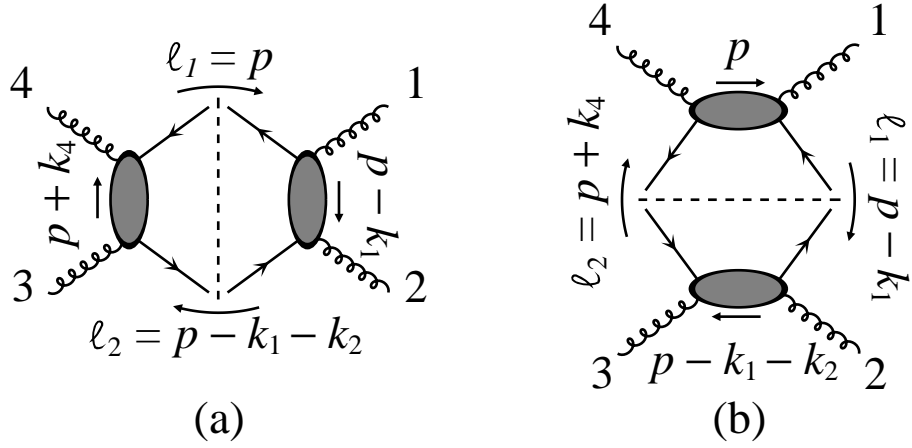,clip=}
\end{center}
\caption[]{
\label{FigTheSCut}
The $s$- and $t$-channel cuts of a one-loop four-gluon
amplitude. The cut lines represent either a fermion or scalar.}
\end{figure}

Consider the $s$-channel ($s = (k_1 + k_2)^2$) cut of the
four-gluon amplitude pictorially represented in \fig{FigTheSCut}a, and
given by the phase-space integral
$$
\eqalign{
-i\, {\rm Disc} \;
A_{4} (1, 2, 3, 4)\Bigr|_{s\rm -cut} & = \int {d^4 p \over (2\pi)^4} \; 
 2\pi \delta^{\tiny(+)}(\ell_1^2 - m^2)\, 2\pi\delta^{\tiny(+)}
 (\ell_2^2 - m^2) \cr
& \hskip 3 cm \times
        A^{\rm tree}_4 (-\ell_1, 1,2,\ell_2) \, 
        A^{\rm tree}_4 (-\ell_2 ,3,4 ,\ell_1) \, ,  \cr}
\equn
\label{TreeProduct}
$$
where $\ell_1=p$ and $\ell_2 = p - k_1 - k_2$, $m$ is the mass of
the particle in the loop, $\delta^{\tiny(+)}$ is 
the positive energy branch of the delta-function
and `Disc' means the discontinuity across the branch cut.
Note that, in performing the sewing of the tree amplitudes, we have
maintained the clockwise ordering of the legs, as required by the
color decomposition.

Since we wish to reconstruct the full amplitude and not just the
imaginary (or absorptive) 
part, it is convenient to replace the phase-space integral
with the cut of an unrestricted loop momentum integral
\cite{SusyFour}.  This is accomplished by replacing the
$\delta$-functions, which impose on-shellness, with propagators:
we replace \eqn{TreeProduct} with
$$
\eqalign{
A_{4} (1, 2, 3, 4)\Bigr|_{s\rm -cut} & = \int 
 {d^4p\over (2\pi)^4} \;    {i\over \ell_1^2 - m^2} \,  
 A^{\rm tree}_4 (-\ell_1, 1,2,\ell_2) \,{i\over \ell_2^2 - m^2} \, 
        A^{\rm tree}_4 (-\ell_2 ,3,4,\ell_1) \cr} \Bigr|_{s\rm -cut} \, . 
\equn
\label{TreeProductDef}
$$
As indicated this equation is valid only on the $s$-cut. A similar 
equation holds on the $t$-cut (where $t = (k_2 + k_3)^2$),
depicted in \fig{FigTheSCut}b.
Up to potential ambiguities in rational functions, one obtains the full 
amplitude by combining the two cuts into a single function.

For example, consider
the amplitude $A_{4} (1^+, 2^+, 3^+,4^+)$ with a massive scalar%
\footnote{In this paper use the the conventional normalization that
the number of physical states in a fundamental representation complex
scalar is two.  In refs.\cite{AllPlus,SusyOne,SusyFour,Fermion} a
supersymmetry preserving normalization with four states in a complex
scalar was used.}
in the loop.  This amplitude is useful for illustrative purposes
because it is particularly simple.  The tree amplitudes entering the
two sides are easily computed from Feynman diagrams and are given by
$$
\eqalign{
A_4^{\rm tree} (-\ell_1, 1^+,2^+,\ell_2) & = 
i m^2 {\spb1.2 \over \spa1.2  ((\ell_1 - k_1)^2 - m^2)} \, , \cr
A_4^{\rm tree} (-\ell_2, 3^+,4^+,\ell_1) & = 
i m^2 {\spb3.4 \over \spa3.4  ((\ell_2 - k_3)^2 - m^2)} \, , \cr}
\equn
$$
where legs $\ell_1$ and $\ell_2$ represent the cut scalar lines.  From
\eqn{TreeProductDef} the scalar loop contribution to the four-gluon
amplitude is
$$
\eqalign{
A_{4}&(1^+, 2^+, 3^+, 4^+)\Bigr|_{s\rm -cut}
\cr
& = m^4 {\spb1.2 \spb3.4 \over \spa1.2 \spa3.4 }
\int {d^{4} p \over (2\pi)^{4}} \; 
        {1\over (p^2 - m^2) ((p - k_1)^2 - m^2) 
            ((p-k_1-k_2)^2 - m^2) ((p + k_4)^2 - m^2)}
         \Bigr|_{s\rm -cut}
\cr
& = i {m^4\over 16\pi^2}
    {\spb1.2 \spb3.4 \over \spa1.2 \spa3.4 } I_4^{D=4}
    \Bigr|_{s\rm -cut}\, , \cr}
\equn\label{TheBadSCut}
$$
where $I_4^{D=4}$ is the scalar integral box function given%
\footnote{The integral given in \eqn{BoxDef} is actually dimensionally
regularized, but there is no difference through $\Ord(\eps^0)$.}
in \eqn{BoxDef} of \app{IntegralsAppendix}. 
The function $I_4^{D=4}$ contains branch cuts since it contains
logarithms and dilogarithms. At this stage the equality of both sides
of the equation is only for the cut in the $s$-channel.

The $t$-channel cut is similar and may be obtained by relabeling the
legs by $1\leftrightarrow 3$.  Using the identity
$$
{\spb2.3 \spb4.1 \over \spa2.3 \spa4.1 } = 
{\spb1.2 \spb3.4 \over \spa1.2 \spa3.4 } \, , 
\equn
$$
the full amplitude is found to have the form
$$
A_{4}(1^+, 2^+, 3^+, 4^+) 
= {i m^4\over 16 \pi^2}
{\spb1.2 \spb3.4 \over \spa1.2 \spa3.4 } I_4^{D=4} + 
\hbox{rational}\, ,
\equn\label{AllPlusMissing}
$$
since it has the correct branch cuts.  The rational terms
have no cuts.  Indeed, if we compare this to the correct
result for the amplitude
$$
\eqalign{
A_{4}(1^+, 2^+, 3^+, 4^+) & 
 = {i \over 16 \pi^2}
{\spb1.2 \spb3.4 \over \spa1.2 \spa3.4 } 
\Bigl(m^4 I_4^{D=4} - {1\over 6} \Bigr) } \, , 
\equn\label{AllPlusCorrect}
$$
we explicitly see that the last term was missed by our naive
application of the Cutkosky rules. We note that setting $m=0$ in 
\eqn{AllPlusCorrect} gives the correct massless scalar loop result \cite{Long}.

\section{Fixing the ambiguity}
\label{FixAmbiguitySection}

We now demonstrate that the last term in \eqn{AllPlusCorrect}, which
is cut-free, may be obtained from the first one with little additional
work but with a more careful application of unitarity.

\subsection{Cuts with dimensional regularization}

The Cutkosky rules are valid in arbitrary dimensions---that is they
may be used with dimensional regularization.  By working to all orders
in the dimensional regularization parameter, $\eps=(4-D)/2$, the
number of potential ambiguities is greatly reduced.  In a massless
theory, it is clear that there are no polynomial ambiguities in
dimensionally regularized expressions.  All terms necessarily must
contain cuts since the dimensions are shifted by $-2\eps$ and must
therefore be proportional to factors of the form $(-K^2)^{-\eps}$,
where $K^2$ is some kinematic variable.  For example, in the
massless case, by working to higher order in $\eps$, the $1/6$ in
\eqn{AllPlusCorrect} appears as
$$
-{(-s)^{1-\eps} + (-t)^{1-\eps} \over 6 (s+t)} + \cdots = {1\over6} 
+ \Ord(\eps)\, , 
\equn
$$
which contains cuts at $\Ord(\eps)$ linked to the $1/6$.  Indeed, this
observation may be used to compute the rational part of massless
amplitudes directly from the cuts \cite{TwoLoopUnitarity,Unpublished}.

For massive theories this simple argument requires modification,
because of the presence of additional dimensionful parameters (the
masses).
A factor of $m^{-2\eps}$ can in principle soak up dimensions to
generate terms without cuts.  We will return to this issue at the 
end of this section.

As a first step in removing the rational function ambiguities, we
modify the calculation of the previous section to be valid to all
orders in the dimensional regularization parameter, $\eps$; as we shall
see this automatically removes most ambiguities. We consider again the
cut integral in \eqn{TheBadSCut} and \fig{FigTheSCut}a, except this
time we replace the four-dimensional loop momenta, $\ell_1$ and
$\ell_2$, with $(4-2\eps)$-dimensional ones.  A simple way to implement
this is to modify the on-shell condition on the cuts to $\ell_1^2
-\mu^2 = m^2$ and $\ell_2^2 - \mu^2 = m^2$, where $\ell_1$ and
$\ell_2$ are left in four-dimensions, but $\mu^\alpha$ is a vector
representing the $(-2\eps)$-dimensional part of the loop
momentum. (See Appendices \ref{HigherDimensionAppendix} and
\ref{MuSlashEtc}.)  With this notation, it is easy to see that the
dimensionally regularized forms of the tree amplitudes are obtained
from \eqn{TheBadSCut} by shifting $m^2 \rightarrow m^2 + \mu^2$.

Thus, the dimensionally regularized expression for the
$s$-channel cut replacing \eqn{TheBadSCut} is
$$
\eqalign{
A_{4}^{\rm scalar} (1^+, 2^+, 3^+, 4^+)\Bigr|_{s\rm -cut}
& = {\spb1.2 \spb3.4 \over \spa1.2 \spa3.4 }
    \int {d^4 p \over (2\pi)^4} 
       {d^{-2\eps} \mu \over (2\pi)^{-2\eps}}\; 
 {(m^2 + \mu^2)^2 \over \D_4}
         \biggr|_{s\rm -cut} \, ,  \cr }
\equn\label{GoodSCut}
$$
where we have explicitly separated out the integration over the
$(-2\eps)$-dimensions. The box denominator is
$$
\D_4 \equiv (p^2 -\mu^2 - m^2) ((p - k_1)^2 -\mu^2 - m^2) 
            ((p-k_1-k_2)^2  -\mu^2 - m^2) 
            ((p + k_4)^2 -\mu^2 - m^2)  \, . 
\equn
$$
In obtaining this form we used the standard prescription that a
four-dimensional vector is orthogonal to a $(-2\eps)$-dimensional
one.  This representation of dimensional regularization follows the
one used in its original derivation \cite{DimensionalRegularization}
and has recently been used by Mahlon in his recursive approach
\cite{Mahlon}.

As discussed in \app{HigherDimensionAppendix}, 
integrals with powers of $\mu$ in the numerator may be expressed in terms
of higher dimension loop integrals, 
$$
\eqalign{
  &
 I_n[(\mu^2)^r] \equiv i (-1)^{n+1} (4\pi)^{2-\eps} 
  \int 
  {d^4 p \over (2\pi)^4} 
  {d^{-2\eps} \mu \over (2\pi)^{-2\eps}} \,
  {(\mu^2)^r \over \D_n}
  =  -\eps (1 - \eps) (2-\eps)\cdots (r-1-\eps)\, 
          I_n^{D=2r+4-2\eps}\, , \cr}
\equn
$$
where $I_n^{D=2r+4-2\eps}$ is the scalar loop $n$-point integral
function evaluated in ($2r+4-2\eps$)-dimensions and $\D_n$ is the
appropriate product of denominators. Thus, the correct $s$-channel cut
to all orders in $\eps$ is
$$
\eqalign{
A_{4}^{\rm scalar} (1^+, 2^+, 3^+, 4^+)\Bigr|_{s\rm -cut}
& =   {i\over (4 \pi)^{2-\eps}} {\spb1.2 \spb3.4 \over \spa1.2 \spa3.4 }
      \K_4 \Bigr|_{s\rm -cut} \, ,\cr }
\equn
$$
where (see \app{KandJ})
$$
\K_4 \equiv I_4[(m^2 + \mu^2)^2] =
m^4 I_4^{D=4-2\eps} - 2 m^2 \eps \, I_4^{D=6-2\eps} - \eps (1-\eps)
         \,  I_4^{D=8-2\eps} \; . 
\equn\label{KFour}
$$

Once again the $t$-channel cut may be obtained by relabeling 
$k_1 \leftrightarrow k_3$, so that the function
$$
A_{4}^{\rm scalar} (1^+, 2^+, 3^+, 4^+)
=  {i\over (4 \pi)^{2-\eps}} {\spb1.2 \spb3.4 \over \spa1.2 \spa3.4 } \, \K_4
 + f(m^2) \, , 
\equn\label{AllPlusAlmostFinal}
$$
has the correct cuts in both the $s$ and $t$ channels.  The function
$f(m^2)$ has no branch cuts in any kinematic channel; it has a form
$\sim \! m^{-2\eps}$, but may contain rational functions of kinematics
variables.

\subsection{Fixing the remaining ambiguity}

In \app{FixingRemainingAppendix} we show that, for the four-gluon
amplitudes considered in this paper, the following procedure uniquely
fixes all cut-free functions.

\begin{enumerate}

\item Add in the tadpole function $I_1$ with a coefficient adjusted so
that the quadratic ultraviolet divergence vanishes. (The quadratic
divergences cancel in gluon amplitudes as discussed, for example, in
ref.~\cite{PeskinSchroeder}.)  This may be implemented by dropping all
explicit $I_1$ functions, then substituting $ J_2 \rightarrow J_2 + I_1 $ and
$I_2^{D=6-2\eps} \rightarrow I_2^{D=6-2\eps} - I_1 $. These
integral functions are defined in \eqns{(\ref{Tadpole}),
(\ref{HigherToLower}) and (\ref{JNoEps})}.

\item Add in the bubble function $I_2(0)$ with a coefficient adjusted so that
the ultraviolet divergence of the amplitude agrees with the well known
one.  
The ultraviolet divergences are 
$$
A_{n}^{\rm loop} \Bigr|_{\rm UV\ singular} =
{1 \over (4\pi)^{2-\eps}}  {\Gamma(1+\eps)\over\eps} 
  (n-2) \Bigl({11 N_c\over 6} -{ n_{\! f} \over 3} 
  - {n_s\over 12}  \Bigr) \, A^{\rm tree}_n \,, 
\equn\label{UVSingularity}
$$
for a theory with $N_c$ colors, $n_{\! f}$ fermion flavors and $n_s$
complex scalars.  The first contribution in the parenthesis is from a
gluon loop and the remaining two from fermion and scalar loops.  
\end{enumerate}
We will use these rules for fixing all cut-free functions in the
amplitudes presented in this paper.

As an alternative to the second step, we may use the known infrared
behavior as the masses vanish.  For massless QCD, the dimensionally
regularized infrared divergences have been categorized in
refs.~\cite{IRSingular}.  This may be used to fix the
coefficients of all $\ln(m_i)$ terms as $m_i\rightarrow 0$.  In this
limit the matching condition for infrared singularities is
$$
\ln(m_i^2) \rightarrow {1\over \eps} + \cdots
\equn
$$
In general, the infrared properties are more powerful since
we can use them to fix coefficients when there are multiple masses
present in an amplitude.

For the amplitudes considered here, we find that the second step is
equivalent to adding in the cut-free self-energy diagrams on external
legs.  For each external gluon leg the bubble contribution is $ A^{\rm
  tree}\, \F/2$, where
$$
{\F}_{\rm scalar} = - {1\over (4\pi)^{2-\eps}} {\Gamma(1+\eps) \over 
 6\eps} \, m^{-2\eps} 
=  -{1\over 6}\,{1\over (4\pi)^{2-\eps}} \, I_2(0) \, , 
\equn
$$
and
$$
{\F}_{\rm fermion} =  -{1\over(4\pi)^{2-\eps}} 
{2\Gamma(1+\eps) \over  3\eps} \, m^{-2\eps} 
=  -{2\over 3}\,{1\over (4\pi)^{2-\eps}} \, I_2(0) \, , 
\equn
$$
corresponding to scalar and fermion loop contributions.

For the $A_4^{\rm scalar}(1^+, 2^+, 3^+, 4^+)$, the two steps are
trivial because $A^{\rm tree}$ vanishes and there are no ultraviolet
divergences.  This means that $f(m^2) = 0$ in 
\eqn{AllPlusAlmostFinal} so that 
$$
A_{4}^{\rm scalar} (1^+, 2^+, 3^+, 4^+)
= {i\over (4 \pi)^{2-\eps}} {\spb1.2 \spb3.4 \over \spa1.2 \spa3.4 } \K_4
\equn\label{AllPlusAnswer}
$$
is the complete amplitude valid to all orders in $\eps$.

The last two terms in $\K_4$, defined in \eqn{KFour}, are proportional
to $\eps$ and will contribute through $\Ord(\eps^0)$ only if the
integrals contain an $\eps^{-1}$.  The integral $I_4^{D=6-2\eps}$ is
finite, but $I_4^{D=8-2\eps}$ has an ultraviolet divergence.
Extracting this divergence either by direct loop integration, or by
use of the integral recursion relation (\ref{HigherToLower}), we have
$$
-\eps(1-\eps) I_4^{D=8-2\eps} =
- {1\over 6} + \Ord(\eps) \, ,
\equn\label{PolyTerm}
$$
which gives the missing rational term in \eqn{AllPlusMissing}.  Hence,
\eqn{AllPlusAnswer} agrees with the correct result
(\ref{AllPlusCorrect}) and contains all rational functions through
$\Ord(\eps^0)$.  Since the ultraviolet divergences of the integral
functions are independent of masses, this rational contribution is
identical to the contribution in the massless case, through
$\Ord(\eps^0)$.  This feature holds for all the amplitudes presented
in this paper.

\section{Fermion loop amplitudes}
\label{FermionLoopAmplitudes}

In this section we consider fermion loop amplitudes.  Fermions are
slightly more complicated than scalars as one must deal with Dirac
algebra, a summary of which is presented in \app{MuSlashEtc}. Apart
from this, the techniques used in the previous section for the scalar
loop carry over to fermion calculations.

\subsection{The $A_4(1^+, 2^+, 3^+, 4^+)$ amplitude}

As a first example, we reconsider the all plus helicity amplitude but with
a massive fermion loop instead of a massive scalar loop. In
this case the tree amplitudes appearing on both sides of the $s$-channel 
cut (see \fig{FigTheSCut}a) are
$$
\eqalign{
  A_4^{\rm tree} (-L_1^f, 1^+, 2^+, L_2^f) & = i {\spb1.2 \over \spa1.2}
   { \bra{-\! L_1} \omega_+ ( \s\mu + m ) \ket{-\! L_2}
     \over     (L_1 - k_1)^2 - m^2} \, ,  \cr
  A_4^{\rm tree} (-L_2^f, 3^+, 4^+, L_1^f) & = i {\spb3.4 \over \spa3.4}
   { \bra{-\! L_2} \omega_+ ( \s\mu + m ) \ket{-\! L_1}
     \over     (L_2 - k_3)^2 - m^2} \, ,  \cr}
\equn
$$
where $\omega_\pm = {1\over 2} (1\pm\gamma_5)$, 
capitalized momenta are $(4-2\eps)$-dimensional and lower case
momenta are four-dimensional. (See \app{IntegralsAppendix}.) 
In these tree amplitudes we suppress the distinction between fermion and
anti-fermion spinors: $u(p)$ versus $v(p)$. We find this convenient
since it provides 
a uniform notation for both types of spinors. When the momentum
follows the direction of the fermion arrow it represents a fermion
state. Conversely, when the momentum flows against the arrow we have an
{\em anti}-fermion state.  However, using this convention
an overall sign error is made for the contribution of a $v\bar{v}$ pair;
$\sum u(-p) \bar{u}(-p) = -\s{p}+m = - \sum v(p)
\bar{v}(p)$. Since this is an overall sign, we may easily correct for it by
hand. Alternatively, one may work with both $u$ and $v$ to obtain
identical results. 

The method for obtaining the cuts is identical to that of the scalar
case in \eqn{GoodSCut}, except that instead of $(\mu^2 + m^2)^2$ the
numerator is initially
$$
\eqalign{
 - \bra{-\! L_1} \omega_+ ( \s\mu + m ) & \ket{ -\! L_2}
 \bra{-\! L_2} \omega_+ ( \s\mu + m ) \ket{ -\! L_1} = \cr
& - \tr [( \s{\ell}_1 + \s\mu - m) \omega_+ (\s\mu + m)
       ( \s{\ell}_2 + \s\mu - m) \omega_+ (\s\mu + m) ] \,,  \cr}
\equn
$$
where the explicit overall minus sign compensates for our suppression
of the distinction between fermions and anti-fermions.  Since $\s\mu$
(like $m$) commutes freely (see \eqn{HelicityProjection}) with the
helicity projection operator, $\omega_+$, we find that the
$\s{\ell}_i$ in the above trace cancel. For example the term,
$$
\cdots \omega_+ \, (\s\mu + m) \s{\ell}_1 \, \omega_+ \, \cdots =
\cdots (\s\mu + m) \, \omega_+ \,\s{\ell}_1 \,\omega_+ \,\cdots =
0 \, .
\equn
$$
The trace thus reduces to a combination of $\s\mu$ and $m$ containing
terms. We readily evaluate what remains as,
$$
\tr [  (\s\mu - m) \omega_+ (\s\mu + m)
                (\s\mu - m) \omega_+ (\s\mu + m)]
=\tr [  \omega_+ (\s\mu^2 - m^2)^2] 
         = 2 (\mu^2 + m^2)^2 \, , 
\equn
$$
so that 
$$
A_{4}^{\fermion} (1^+,2^+,3^+,4^+) = - {2i \over (4\pi)^{2-\eps}}
{\spb1.2 \spb3.4 \over \spa1.2 \spa3.4 } \K_4 \, , 
\equn
$$
where $\K_4$ is defined in \eqn{KFour}.  We have fixed the cut-free
function, by following the same steps as for the scalar loop case
(\ref{AllPlusAnswer}).  Thus for this helicity, the contribution for a
Dirac fermion is minus twice that of a complex scalar given in
\eqn{AllPlusAnswer}. This is the correct result as may be verified
from a supersymmetry identity \cite{Susy}.  In summary, for this
helicity, we have obtained the correct result for a massive quark
loop, with little more work than with a naive application of
unitarity.

\subsection{The $A_4(1^-,2^-,3^+, 4^+)$ amplitude}

Having calculated the simplest of the closed fermion loop
sub-amplitudes consider now one of the more difficult ones. The
$A_{4}(1^-,2^-,3^+,4^+)$ sub-amplitude does not have an
$s\leftrightarrow t$ exchange symmetry. Accordingly, we must explicitly 
calculate both cuts and combine the information to
obtain the complete result. Also, this amplitude contains
non-vanishing cut-free
functions whose coefficients will be fixed from the ultraviolet
divergences.

The easier cut is in the $s$-channel, so we calculate it
first. The required tree-level amplitudes are,
$$
\eqalign{
  A_4^{\rm tree} (-L_1^f, 1^-, 2^-, L_2^f) & = i {\spa1.2 \over \spb1.2}
   { \bra{-\! L_1} \omega_- ( \s\mu + m ) \ket{-\! L_2}
     \over     (L_1 - k_1)^2 - m^2} \, ,  \cr
  A_4^{\rm tree} (-L_2^f, 3^+, 4^+, L_1^f) & = i {\spb3.4 \over \spa3.4}
   { \bra{-\! L_2} \omega_+ ( \s\mu + m ) \ket{-\! L_1}
     \over     (L_2 - k_3)^2 - m^2} \, . \cr}
\equn
$$
As before, these combine to give a contribution to the $s$-cut,
$$
\eqalign{
A_{4} (1^-, 2^-, 3^+, 4^+)\Bigr|_{s\rm -cut}
& = -i {t\over s} A_4^{\rm tree} 
    \int {d^4 p \over (2\pi)^4} 
       {d^{-2\eps} \mu \over (2\pi)^{-2\eps}}\cr
& \hskip 2.5 cm \times
  {  \bra{-\! L_1} \omega_- ( \s\mu + m ) \ket{-\! L_2}
  \bra{-\! L_2} \omega_+ ( \s\mu + m ) \ket{-\! L_1}   \over \D_4}
        \biggr|_{s\rm -cut} \,,  \cr }
\equn
$$
where $\ell_1 = p$, $\ell_2 = p - k_1 - k_2$ and 
we use (see ref.~\cite{ManganoReview})
$$
{\spa1.2 \spb3.4 \over \spb1.2 \spa3.4 }
= - {t \over s}\, {\spa1.2^4 \over \spa1.2\spa2.3\spa3.4\spa4.1}
= i\, {t \over s} \, A_4^{\rm tree}(1^-, 2^-, 3^+, 4^+) \, . 
\equn\label{TreeMMPP}
$$

Concentrating on the numerator Dirac algebra under the integral, we
create a trace,
$$
\eqalign{
  \bra{-\! L_1} \omega_- ( \s\mu + m ) \ket{-\! L_2} &
  \bra{-\! L_2} \omega_+ ( \s\mu + m ) \ket{-\! L_1}
        = \cr
  & \tr [ (\s{\ell}_1+\s\mu - m) \omega_- ( \s\mu + m )
        (\s{\ell}_2 +\s\mu-m) \omega_+ ( \s\mu + m )]  \, . \cr}
\equn
$$
Unlike the all plus case, the helicity projection operators do not
cancel the four-dimensional momenta. Instead they
cancel the $(\s\mu-m)$ contributions from the spinors on the cut.
The above trace reduces to,
$$
\eqalign{
 \tr [\omega_+ \s{\ell}_1 ( \s\mu + m )
        \s{\ell}_2  ( \s\mu + m )]
& =   ( \mu^2 + m^2 ) \tr [ \omega_+ \s{\ell}_1 \s{\ell}_2]
 =  ( \mu^2 + m^2 ) \,  2 \ell_1 \c \ell_2 \cr
& = 2 (\mu^2 + m^2)^2 -  s (\mu^2 + m^2) \, , \cr} 
\equn
$$
where we have used the fact that on the cut $\ell_1^2 = \ell_2^2 =
\mu^2 + m^2$.  This integrand exhibits a supersymmetry decomposition
as the first term is identical (up to an overall factor of $-2$) to
the contribution of complex scalar loop.

Performing the integration,
the $s$-cut of the amplitude becomes
$$
A_4^{\fermion} (1^-, 2^-, 3^+, 4^+)\Bigr|_{s\rm -cut}
=  - {1 \over (4\pi)^{2-\eps}}
A_4^{\rm tree} \Bigl(
- 2 {t\over s}\, \K_4 + t\, \J_4
\Bigr)\Bigr|_{s\rm -cut} \, , 
\equn\label{MMPPSCut}
$$
where $\J_4$ and $\K_4$ are defined in 
\eqns{(\ref{DefineJ}) and (\ref{DefineK})}.  

The $t$-cut, depicted in \fig{FigTheSCut}b, is more complicated. It
may be constructed from the following two tree amplitudes,
$$
\eqalign{
  A_4^{\rm tree} (-L_1^f, 2^-, 3^+, L_2^f) & =
        i\, \pol_2^- \cdot \ell_1 \,
        { \bra{-\! L_1} \s{\pol}^+_3 \ket{-\! L_2}
        \over
        (\ell_1 - k_2)^2 - \mu^2 - m^2  } \, , \cr
  A_4^{\rm tree} (-L_2^f, 4^+, 1^-, L_1^f) & =
       i \, \pol_4^+ \cdot \ell_2 \, 
        { \bra{-\! L_2} \s{\pol}^-_1 \ket{-\! L_1}
        \over (\ell_2 -k_4)^2 -\mu^2 - m^2} \, , }
\equn\label{TCutTrees}
$$
where the reference momenta $q_i$ for the polarization vectors
$\pol_i$ are are $q_2 = k_3$, $q_3 = k_2$, $q_4 = k_1$, 
and $q_1 = k_4$.  This is a particularly good choice of reference momenta 
as each tree amplitude is described by a single Feynman diagram.
As depicted in \fig{FigTheSCut}b, for the $t$-channel
the cut momenta are
$$
\ell_1 = p - k_1 \,, \hskip 2 cm 
\ell_2 = p + k_4 \,.
\equn
$$
This representation of the tree amplitudes may be directly obtained
from Feynman diagrams after using the spinor helicity form of the
polarization vectors.  (As discussed in \app{MuSlashEtc}, identity
(\ref{PolSlash}) may be used on the $\s\pol$ only when it is adjacent
to a four-dimensional quantity.)

Using \eqn{TreeProduct} and combining the two tree amplitudes into a
trace, the $t$-cut becomes,
$$
\eqalign{
A_{4}^{\rm fermion} (1^-, 2^-, 3^+, 4^+)\Bigr|_{t\rm -cut}
& = - \int {d^4 p \over (2\pi)^4} 
       {d^{-2\eps} \mu \over (2\pi)^{-2\eps}} \cr
& \hskip 1.5 cm \times { \pol_2 \c \ell_1 \pol_4\c \ell_2 
  \tr[(-\s{\ell}_1 -\! \s\mu +\! m) \s\pol_3
         (-\s{\ell}_2 -\! \s\mu +\! m) \s\pol_1]  \over \D_4}
        \biggr|_{t\rm -cut} . \cr
}
\equn\label{UglyTCut}
$$
The numerator may be rewritten as 
$$
\eqalign{ {\cal N} & =
  - \pol_2 \c \ell_1 \pol_4 \c \ell_2 \tr
  [(\s{\ell}_1 +\! \s\mu -\! m) \s\pol_3 (\s{\ell}_2 +\! \s\mu -\! m)
  \s\pol_1] \cr
  & = - 2 {\cal N}_1 - {\cal N}_2 \,, \cr}
\equn\label{TCutNumerator}
$$
where 
$$
\eqalign{
 {\cal N}_1 & =  4 \pol_2\cdot \ell_1 \, \pol_4 \cdot \ell_2 
  \, \pol_3\c\ell_1 \, \pol_1 \c\ell_2 \, , \hskip 1.5 cm 
 {\cal N}_2  =  2 t \, \pol_2\cdot \ell_1
  \, \pol_4 \cdot \ell_2 \,\pol_1 \c \pol_3 \, .\cr}
\equn
$$
We have dropped terms with an odd power of $\mu^\alpha$ (see
\app{IntegralsAppendix} for the relevant discussion).  ${\cal N}_1$
may be recognized as the numerator of a cut complex
scalar loop, while the second is the additional piece for a
fermion loop.  Thus, this integrand again obeys a supersymmetry
decomposition.

To evaluate the amplitudes, we use spinor helicity.  From
\eqn{PolarizationVector}, and the choice of reference momenta given
below \eqn{TCutTrees}, the polarization vectors satisfy
$$
\eqalign{
& \pol_1^- \c \pol_3^+ = {\spa1.2 \spb3.4 \over \spa2.3 \spb1.4} \,,
  \hskip 1 cm
  \pol_1^- \c \ell_2 = {\sandpp4.{\s{\ell}_1}.1 \over \sqrt{2} \spb1.4 } \,,
  \hskip 1 cm
  \pol_2^- \c \ell_1 = {\sandpp3.{\s{\ell}_1}.2 \over \sqrt{2} \spb2.3
    } \,, \cr
  & \hskip 2.5 cm
  \pol_3^+ \c \ell_1 = {\sandmm2.{\s{\ell}_1}.3 \over \sqrt{2} \spa2.3 } \, , 
  \hskip 1 cm
  \pol_4^+ \c \ell_2 = {\sandmm1.{\s{\ell}_1}.4 \over \sqrt{2} \spa1.4 } \, .
      \cr}
\equn
$$
It is convenient to evaluate separately the two numerator terms
of~\eqn{TCutNumerator}. The first corresponds, up to a factor
of $-2$, to the contribution of a complex scalar loop, whose numerator
is 
$$
\eqalign{
  {\cal N}_1  & =
  {\sandpp4.{\s{\ell}_1}.1 \over \spb1.4 }
  {\sandpp3.{\s{\ell}_1}.2 \over \spb2.3 }
  {\sandmm2.{\s{\ell}_1}.3 \over \spa2.3 }
  {\sandmm1.{\s{\ell}_1}.4 \over \spa1.4 } \cr
  & = {1\over t^2} {\sandmm2.{\lsl_1}.3}^2 {\sandpp4.{\lsl_1}.1}^2 \,.\cr}
\equn
$$

By multiplying and dividing by $\spa3.4^2 \spb1.2^2$, ${\cal N}_1$ can
be rewritten in terms of a trace as
$$
{
  {\cal N}_1 = {1\over t^2}
  {\sandmm2.{\lsl_1}.3}^2 { \spa3.4^2 \over \spa3.4^2 }
  {\sandpp4.{\lsl_1}.1}^2 { \spb1.2^2 \over \spb1.2^2 }
  =
  {i \over s^3 t}\, A_4^{\rm tree}(1^-, 2^-, 3^+, 4^+)
  \tr_+[\ell 3 4 \ell 1 2 \ell 3 4 \ell 1 2] } \, , 
\equn
$$
where we have defined $\ell_1 \equiv \ell$, and for brevity 
represent the external momenta by their indices and neglect
slashing the momenta inside the trace. (We use the notation,
$\tr_\pm(\ldots) = \tr(\omega_\pm \ldots)$.) Since tensor box
integrals are generally difficult to evaluate (without the aid of a
computer), we will avoid such terms.  To simplify the trace, we
commute the $\ell$s within the trace towards each other. This
generates more simple integrals, with inverse propagators or fewer
powers of loop momentum in the numerator:
$$
\eqalign{
\tr_+[\ell 3 4 \ell 1 2 \ell 3 4 \ell 1 2]
& =
 \tr_+[\ell 3 4 \ell 1 2 \ell 3 4 2] \lor{\ell}.1
-\tr_+[\ell 3 4 \ell 1 2 \ell 3 4 1] \lor{\ell}.2
+\tr_+[2 3 4 \ell 1 2 \ell 3 4 1 ] (\mu^2 +m^2) \cr
& =
 \tr_+[\ell 3 4 \ell 1 2 \ell 3 4 2] \lor{\ell}.1
-\tr_+[\ell 3 4 \ell 1 2 \ell 3 4 1] \lor{\ell}.2
+ \tr_+[2 3 4 2 \ell 3 4 1 ] (\mu^2+ m^2) \lor{\ell}.1 \cr
& \hskip 1 cm
- \tr_+[2 3 4 1  \ell 3 4 1 ] (\mu^2 + m^2) \lor{\ell}.2
+ \tr_+[2 3 4 1 2 3 4 1 ] (\mu^2 + m^2)^2  \, . \cr }
\equn
$$
Using
$$
\lor{\ell}.1 \equiv 2 \, \ell \cdot k_1 =  p^2 - \mu^2 -m^2  \, , 
\hskip 1.5 cm 
\lor{\ell}.2 \equiv 2 \, \ell \cdot k_2 = -(p - k_1 -k_2)^2 
 +\mu^2 + m^2  \,,
\equn
$$
this becomes
$$
\eqalign{
&
\tr_+[\ell 3 4 \ell 1 2 \ell 3 4 2] ( p^2 - \mu^2 -m^2)
+\tr_+[\ell 3 4 \ell 1 2 \ell 3 4 1]  ((p - k_1 -k_2)^2 -\mu^2 - m^2 )
\cr
& 
+ \tr_+[2 3 4 2 \ell 3 4 1 ] (\mu^2+ m^2) (p^2 - \mu^2 -m^2)
+ \tr_+[2 3 4 1  \ell 3 4 1 ] (\mu^2 + m^2)  ((p - k_1 -k_2)^2 -\mu^2 - m^2)
\cr
& 
+ \tr_+[2 3 4 1 2 3 4 1 ] (\mu^2 + m^2)^2 \,,   }
\equn
\label{TraceMess}
$$
which exhibits inverse propagators in the numerator.  After cancelling
these inverse propagators, against those present in \eqn{UglyTCut}, we
obtain tensor triangle and scalar box integrals.

One way to evaluate the tensor integrals is via Feynman
parameterization.  A convenient way to do this is to have a uniform
labeling of Feynman parameters for all diagrams: for the triangles,
the Feynman parameter corresponding to the missing internal propagator
is set to zero.  With this convention, the Feynman parameter shift 
for all diagrams is,
$$
\ell = p - k_1 = q - k_1 a_1 + k_2 (a_3 + a_4) + k_3 a_4 \,,
\equn \label{ParamShift}
$$
where $q$ is the shifted loop momentum.  For example,
the first term in \eqn{TraceMess} corresponds to the triangle integral
$$
{i \over s^3 t}\, A_4^{\rm tree}
\int {d^4 p \over (2\pi)^4} 
       {d^{-2\eps} \mu \over (2\pi)^{-2\eps}}\;  
{\tr_+[\ell 3 4 \ell 1 2 \ell 3 4 2]   \over 
 ((p - k_1)^2 -\mu^2 - m^2) ((p-k_1-k_2)^2  -\mu^2 - m^2) 
            ((p + k_4)^2 -\mu^2 - m^2)  } \, . 
\equn
$$
Performing the Feynman parameter shift in \eqn{ParamShift}, but with
$a_1 = 0$ since the corresponding propagator cancels, and integrating
over $q$ and $\mu$ we obtain (see \app{FeynmanParamAppendix}) 
$$
\eqalign{
{2 i \over s^3 t}\, A_4^{\rm tree} 
 \int_0^1 d^3 a_i \; \delta(1 - {\textstyle \sum_j} a_j)
& 
\int {d^4 q \over (2\pi)^4} 
       {d^{-2\eps} \mu \over (2\pi)^{-2\eps}}\;  
{\tr_+[q  3 4 (q + k_2 a_3) 1 2 q 3 4 2]   \over 
(q^2  - a_2 a_4 s_{23} - m^2 -\mu^2)^3 } \cr 
& =  - {1\over (4\pi)^{2-\eps}}  A_4^{\rm tree} {1\over 2 \, s^3 t}\,
       \tr_+ [ \gamma_\nu 3 4 2 \gamma^\nu 3 4 2] \,
                s I_3^{(1),\, D=6-2\eps}[a_3] \, \cr
& = 0 \, , \cr}
\equn
$$
where we use $\s{k}_i \s{k}_i = 0$, momentum conservation and
$$
\gamma_\nu \ksl_m \ksl_n \gamma^\nu = 4 k_m \cdot k_n \, ,  \hskip 1 cm 
\gamma_\nu \ksl_m \ksl_n \ksl_l \gamma^\nu  = - 2 \ksl_l \ksl_n \ksl_m \, .
\equn
$$
(Note that the Lorentz indices of the $\gamma$-matrices are
four-dimensional.  This means that the required $\gamma$-matrix
identities are the standard four-dimensional ones.)  As defined in
\app{FeynmanParamAppendix}, the integral function $I_n[a_i]$ is a
Feynman parameterized form of the loop integral with a factor of $a_i$
in the numerator. The superscript on $I_3^{(1)}$ indicates the triangle 
obtained from the box by removing the propagator between legs 4 and 1,
as discussed in \app{IntegralsAppendix}.

Similarly, evaluating all the terms in \eqn{TraceMess} we obtain
$$
\eqalign{
  A_4^{\scalar} (1^-, 2^-,& 3^+, 4^+) \Bigr|_{t\rm - cut}
  = {1 \over (4\pi)^{2-\eps}} A_4^{\rm tree} {1 \over s^3 t}
  \biggl( 
  - \half \tr_+ [ 2 3 4 \gamma_\nu 1 2 \gamma^\nu 3 4 1 ]
                I_3^{(3),D=6-2\eps}[a_4] \cr
& \hskip 4 cm 
  + \tr_+ [ 2 3 4 1 2 3 4 1 ] \J_3^{(3)} [a_4]
  - \tr_+ [ 2 3 4 1 2 3 4 1 ] \K_4
\biggr) \, ,\cr} 
\equn
$$
where $
\J_3^{(3)} [a_4] \equiv m^2 I_3^{(3)}[a_4] -
\eps I_3^{(3), \, D=6-2\eps}[a_4]$. (cf. \eqn{DefineJ}.)

Using momentum conservation to evaluate the traces we have
$$
\eqalign{
  A_4^{\scalar} & = 
  {1 \over (4\pi)^{2-\eps}} A_4^{\rm tree} \Bigl(
  I_3^{(3), D=6-2\eps}[a_4]
  + {t\over s} J_3^{(3)}[a_4]
  - {t\over s} K_4
  \Bigr) \Bigr|_{t-\rm cut} \, .\cr}
\equn
$$
The $\gamma_5$ in $\tr_+$ vanishes for a four-point function, because
there are only three independent momenta to contract into the
Levi-Civita tensor.

As a final step, we may convert the single parameter triangles to
scalar integrals using the integral relation
(\ref{LinearIntegralRelation}), which explicitly yields
$$
I_3^{(3)}[a_4] = {1\over t} \, I_2^{(1,3)}\, , \hskip 2 cm 
I_3^{(3), D=6-2\eps}[a_4] = {1\over t} \, I_2^{(1,3), D=6-2\eps} \,, 
\equn
$$
where we have dropped those bubbles without a $t$-channel cut.
This gives the final form of the $t$-cut as
$$
  A_4^{\scalar} = 
  {1 \over (4\pi)^{2-\eps}} A_4^{\rm tree} \Bigl(
  {1\over t} I_2^{(1,3), D=6-2\eps}
  + {1 \over s} J_2^{(1,3)}
  - {t\over s} K_4
  \Bigr) \Bigr|_{t-\rm cut} \, .
\equn\label{TCutmmpp}
$$

Combining the $t$- and $s$-cuts in \eqn{TCutmmpp} and minus half of
the first term in \eqn{MMPPSCut}, we find that the full scalar loop
amplitude is of the form,
$$ 
  A_{4}^{\scalar}(1^-, 2^-, 3^+, 4^+) = 
  {1 \over (4\pi)^{2-\eps}} A_4^{\rm tree}
  \Bigl(
  {1 \over t}  \, I_2^{(1,3), D=6-2\eps}
  + {1\over s} \J_2^{(1,3)}
  - {t \over s} \K_4
  \Bigr)
  + f(m^2) \, , 
\equn
$$
where $f(m^2)$ is function of the masses which does not
have a cut.

We still have to evaluate the uniquely fermionic term in
\eqn{TCutNumerator}, ${\cal N}_2$, which combines with the scalar loop
contribution to form the complete fermion loop.  Since it contains
only two powers of loop momentum in the numerator, instead of four, it
is simpler to evaluate than the above scalar loop term.  The
result of such an evaluation is
$$
A_4^{\rm fermion}(1^-, 2^-, 3^+, 4^+) = 
- 2 A_4^{\rm scalar}(1^-, 2^-, 3^+, 4^+)
- {1 \over (4\pi)^{2-\eps}}
     A_4^{\rm tree} \, (t\, \J_4 - I_2^{(1,3)})
     + g(m^2)  \, . 
\equn
$$

Finally, we fix the form of $f(m^2)$ and $g(m^2)$ by correcting the
ultraviolet divergences with the two-step
procedure in \sec{FixAmbiguitySection}.2 to obtain  the final result
$$
\eqalign{
  A_{4}^{\scalar}(1^-, 2^-, 3^+, 4^+) & = 
  {1 \over (4\pi)^{2-\eps}} A_4^{\rm tree}
  \Bigl(
  {1 \over t}  ( I_2^{D=6-2\eps}(t) - I_1 )
  + {1 \over s} ( \J_2(t) + I_1 )
  - {t \over s} \K_4
  - {1 \over 3} I_2(0)
  \Bigr) \,,
  \cr
  A_4^{\rm fermion}(1^-, 2^-, 3^+, 4^+) & = 
  - 2 A_4^{\rm scalar}(1^-, 2^-, 3^+, 4^+)
  - {1 \over (4\pi)^{2-\eps}}
     A_4^{\rm tree} \, (t\, \J_4 - I_2(t) + 2 I_2(0) ) \, , 
  \cr
}
\equn\label{Finalmmpp}
$$
where $A_4^{\rm tree}$ is defined in \eqn{TreeMMPP}.  These
expressions for the amplitudes are valid to all orders in the
dimensional regularization parameter, $\eps$.

These amplitudes are bare ones before subtraction of the
ultraviolet singularities.  To obtain the $\overline{\rm MS}$ renormalized
amplitudes (in the FDH regularization scheme), we subtract the
ultraviolet divergences 
$$
\eqalign{
& - {(n-2)\over 3} {\Gamma(1+\eps) \over (4\pi)^{2-\eps} }
A_n^{\rm tree}\, {1\over \eps}\,,
         \hskip 1 cm \hbox{for fermion loop,}  \cr
& - {(n-2)\over 12} {\Gamma(1+\eps) \over (4\pi)^{2-\eps} }
A_n^{\rm tree}\, {1\over \eps}\,, 
         \hskip 1 cm \hbox{for scalar loop,}   \cr}
\equn\label{UVSubtractions}
$$
where $n=4$ is the number of external legs.

After appropriately redefining the coupling constant
\cite{KunsztFourPoint}, the results in \eqn{Finalmmpp}, are identical
in the 't Hooft-Veltman scheme (where observed particles are in
four-dimensions, but unobserved ones are continued to $D=4-2\eps$).
(The gluon loop contribution is, however, shifted \cite{Long}.)
One may also convert to the conventional dimensional regularization
scheme, by accounting for the external $\eps$-helicities
\cite{EpsHel}.

Whilst unitarity unambiguously fixes the correct sign for the cut of
any loop amplitude, it is frequently convenient to ignore the overall
sign of a cut as it may be determined in the final stages of a
calculation.  By comparing integral functions with cuts in multiple
channels we can usually fix the relative signs of the separately calculated
contributions. The overall sign of the amplitude can be obtained
by comparing with known results in the massless limit, or in the case
of five- or higher-point amplitudes, factorization properties are a
sufficient constraint. Alternatively, these consistency requirements
may be used as checks.

\section{Remaining amplitudes}
\label{RemainingSection}

The remaining four-gluon helicity amplitudes are straightforward to calculate
using the above techniques.  For completeness we quote the results:
$$
\eqalign{
  A_4^{\scalar}(1^-,&\, 2^+, 3^+, 4^+) =
  {i \over (4\pi)^{2-\eps}}
  { \spb2.4^2 \over \spb1.2 \spa2.3 \spa3.4 \spb4.1 }
  { s t \over u } \cr
  \times \biggl(&
      { t(u-s) \over s u } J_3(s)
      + { s(u-t) \over t u } J_3(t)
      - { t-u \over s^2 } (J_2(s)+I_1)
      - { s-u \over t^2 } (J_2(t)+I_1)
      + { s t \over 2 u } J_4
      + K_4
  \biggr) \, ,  \cr
  A_{4}^{\scalar} (1^-,&\, 2^+, 3^-, 4^+) =
  {1 \over (4\pi)^{2-\eps}} A_4^{\rm tree} \biggl(
       { s t(s-t) \over u^3 } J_3(t)
      +{ s t(t-s) \over u^3 } J_3(s)
      -{ t s^2 \over u^3 } I_2(t)
      -{ s t^2 \over u^3 } I_2(s)
      \cr
      &
      -{ s \over t u } (I_2^{D=6-2\eps}(t)-I_1)
      -{ t \over s u } (I_2^{D=6-2\eps}(s)-I_1)
      -{ s \over u^2 } (J_2(t)+I_1)
      -{ t \over u^2 } (J_2(s)+I_1)
      \cr
      &
      -{ s t \over u^2 } I_3^{D=6-2\eps}(t)
      -{ s t \over u^2 } I_3^{D=6-2\eps}(s)
      +{ s^2 t^2 \over u^3 } I_4^{D=6-2\eps}
      -{ s t \over u^2 } K_4
      -{ 1 \over 3 } I_2(0)
  \biggr) \, , 
\cr
}
\equn
$$
where $u = - s -t$ and 
$$
A_4^{\rm tree} (1^-, 2^+, 3^-, 4^+) = i 
{\spa1.3^4 \over \spa1.2\spa2.3\spa3.4\spa4.1} \, . 
\equn
$$
The remaining fermion loop amplitudes are
$$
\eqalign{
A_{4}^{\fermion}(1^-, 2^+, 3^+, 4^+) & 
= - 2 A_{4}^{\scalar}(1^-, 2^+, 3^+, 4^+)\, ,  \cr
A_{4}^{\fermion}(1^-, 2^+, 3^-, 4^+)
& =  -2 A_{4}^\scalar(1^-, 2^+, 3^-, 4^+) \cr
& \hskip .5 cm 
- {1 \over (4\pi)^{2-\eps}} A_4^{\rm tree} \Bigl(
    { t \over u } I_2(s)
  + { s \over u } I_2(t)
  - { st \over u } J_4
  - { st \over u } I_4^{D=6-2\eps}
  + 2 I_2(0) 
\Bigr) \, .
\cr}
\equn
$$
These amplitudes explicitly exhibit the supersymmetry decomposition 
discussed in refs.~\cite{FiveGluon,SusyDecomp,SusyFour}, since
the fermion loop amplitudes contain pieces which are naturally expressed 
in terms of the scalar loop amplitudes.

We have explicitly checked that the four-point amplitudes obtained
via unitarity are in full agreement with a more conventional
diagrammatic calculation.  (The forms obtained via a conventional
Feynman diagram calculation are in general somewhat more complicated
than those obtained via unitarity; the two forms may be compared using
the integral recursion formulas (\ref{HigherToLower})).  As before, to
obtain the renormalized amplitudes from the above bare amplitudes we
subtract from the amplitudes the quantities in \eqn{UVSubtractions}.

The methods we have presented here can also be applied to computing
higher point functions.  For example we have obtained the simplest of
the five-gluon amplitudes:
$$
\eqalign{
& A_5^{\fermion}(1^+, 2^+, 3^+, 4^+, 5^+)
= -2 A_5^{\scalar}(1^+, 2^+, 3^+, 4^+, 5^+) \cr
& \hskip .5 cm 
 =   {2i\over \spa1.2 \spa2.3 \spa3.4 \spa4.5 \spa 5.1}
{1 \over(4\pi)^{2-\eps} }   \cr
&  \hskip 1 cm \times
{1\over \tr_5[1234]} \Bigl[ s_{23} s_{34} \tr_-[5124] \K_4^{(1)}
        + s_{34} s_{45} \tr_-[1235] \K_4^{(2)}
        + s_{45} s_{51} \tr_-[2341] \K_4^{(3)} \cr
& \hskip 2 cm
        + s_{51} s_{12} \tr_-[3452] \K_4^{(4)}
        + s_{12} s_{23} \tr_-[4513] \K_4^{(5)} 
        - s_{12} s_{23} s_{34} s_{45} s_{51} \K_5 \Bigr]\,,  \cr}
\equn\label{FiveGluonAmpl}
$$
where $s_{ij} = (k_i + k_j)^2$, $\tr_5[A] = \tr[\gamma_5 A]$ and
$\tr_-[A] = {1\over 2} \tr[(1-\gamma_5)A]$.  The function $\K_4^{(i)}$
is obtained from \eqns{(\ref{ExtMassBoxDef}), (\ref{HigherToLower})
and (\ref{DefineK})}.  We may use \eqn{PentagonInt} to express $K_5$ in
terms of $\K_4^{(i)}$.  The $(i)$ label on $\K_4$ follows the same
convention as that used for $I_4$.  Extending the discussion in 
\app{FixingRemainingAppendix} to this case, it remains true that 
the only cut free functions that can appear are $I_1$ and $I_2(0)$.
Since the corresponding tree amplitude vanishes, this amplitude must
be infrared and ultraviolet finite; the coefficients of these
two cut-free functions vanish as was the case for \eqn{AllPlusAnswer}.

 A few simple consistency checks may be performed on this amplitude:
in the massless limit through $\Ord(\eps^0)$ it reproduces previously
calculated results \cite{FiveGluon}.  Furthermore, it is not difficult
to check that it satisfies appropriate limits as any two external
momenta become collinear \cite{TreeCollinear,AllPlus,Factorization}.
(In performing the check to all orders in $\eps$, integral identities
of the form discussed in section~5.2 of ref.~\cite{Factorization} are
required.)

In obtaining these results, no computerization was necessary.  For
other helicity structures, or different external particles, the
amplitudes become sufficiently complicated that computer assistance is
useful.


\section{Summary and discussion}
\label{SummarySection}

In general, it is far easier to build amplitudes from their analytic
properties than from diagrams.  In order to use unitarity
\cite{Cutting,PeskinSchroeder} as a computational tool for complete
amplitudes, one must maintain control over all potential cut-free
contributions. A powerful constraint is that the functions entering
into an amplitude are not arbitrary, but must be composed of a
definite set of integral functions. This has recently been used to
obtain certain infinite sequences of one-loop amplitudes
\cite{SusyFour,SusyOne}.  With unitarity, the basic building blocks of
one-loop amplitudes are tree-level amplitudes, which generally are far
more compact than individual diagrams, especially in a helicity
representation.  Unitarity is also esthetically appealing in providing
a way to build further amplitudes using on-shell quantities.

In this paper, we extended the unitarity method of
refs.~\cite{SusyFour,SusyOne} to obtain particular examples of
one-loop amplitudes with a uniform mass in the loop.  We have computed
all four-gluon helicity amplitudes with massive fermion or scalar
loops.  These amplitudes have not previously been computed, and are
useful for investigating the heavy quark thresholds in elastic
glue-glue scattering.  If unexpected structure were to be observed in
the experimental jet cross-sections at heavy quark thresholds, a
detailed analysis of this effect would be warranted.

We have proven that all potential cut-free functions that may appear
in these amplitudes can be fixed with a knowledge of the ultraviolet
or infrared divergences.  By performing the calculation to all orders
in the dimensional regularization parameter, only a few cut-free
functions can appear in the amplitudes.  These cut-free functions are
integrals with no dependence on external kinematic variables. For the
examples treated in this paper with a uniform mass in the loop,
computing to all orders in $\eps$ (where $D=4-2\eps$) takes little
extra work as compared to working through $\Ord(\eps^0)$.  The reason
is that the $\eps$ components of the loop momentum act as a uniform
mass in the loop that is then integrated over
\cite{DimensionalRegularization,Mahlon}.  We have also demonstrated that
the coefficients of the cut-free integral functions may be fixed using
well known ultraviolet or infrared divergences of gauge theories as
`boundary conditions'.

The amplitudes obtained in this paper had a number of generic
features.  Firstly, supersymmetry relations
\cite{FiveGluon,SusyDecomp} between contributions to the amplitudes
were identifiable at the level of integrands.  This may be contrasted
to conventional Feynman diagrams where no such simple identification
is possible. This allowed us to compute the scalar loop contribution
as a byproduct of the fermion result.  A second property was that
rational functions (such as in \eqn{PolyTerm}) appearing in the
massive amplitudes were identical to ones appearing in massless
amplitudes.  An encouraging feature of our explicit calculations is
that intermediate expressions did not grow in comparison to the size
of final results.

Unitarity methods may also be applied to higher-point massive
calculations; as an example we have presented the contribution of a
massive fermion loop to the five-gluon amplitude with identical
helicities.  For general processes one would need to extend the
arguments presented in this paper. This type of investigation of the
cut structure of integral functions for an arbitrary number of
external legs has already been carried out for the massless case
\cite{SusyOne}.

The unitarity methods discussed in this paper may be combined with other
techniques.  For example, for five- and higher-point amplitudes,
factorization provides a strong constraint 
\cite{TreeCollinear,AllPlus,SusyFour,Factorization}. Supersymmetry
\cite{Susy,SusyDecomp} and string motivated \cite{Long,StringBased}
ideas can also be incorporated into the unitarity approach. One should
also be able to devise automated amplitude evaluation programs
\cite{Automated} based on unitarity.

The same types of methods as those discussed in this paper can be
applied to cases where there are a variety of masses in the loop, such
as for heavy fermion production, and will be discussed elsewhere
\cite{Future}.  One should be able to use unitarity as a tool for
constructing two-loop amplitudes \cite{TwoLoopUnitarity}, but to
perform four- or higher-point calculations one would first need to
construct a table of two-loop integrals \cite{TwoLoopIntegrals}.

\section*{Acknowledgments}

We thank L. Dixon, D. Dunbar and D. Kosower for extensive valuable
discussions and collaborations.  We also thank J.B.\ Tausk and G.D.\
Mahlon for discussions at the Aspen Center for Physics, where a part
of this paper was written.  This work was supported by the DOE under
contract DE-FG03-91ER40662 and by the Alfred P. Sloan Foundation under
grant BR-3222.

\appendix

\section{Integrals}
\label{IntegralsAppendix}

In this appendix we collect expressions for the integrals
\cite{Integrals,VNV,IntegralRecursion} we use in this paper.  The
$D$-dimensional loop integrals considered in this paper are defined by
$$
  I_n^{D}[P^{\alpha_1} \cdots P^{\alpha_m}] 
= i (-1)^{n+1} (4\pi)^{D/2} \int \frac{
    d^{D} P }{ (2\pi)^{D} }
  \frac{P^{\alpha_1} \cdots P^{\alpha_m}} {(P^2 - m^2) \ldots (
    (P-\sum_{i=1}^{n-1} k_i )^2 - m^2) } \,,
  \equn\label{GeneralLoop}
$$
where $P$ is a $D$ dimensional momentum and the
external momenta, $k_i$, are four-dimensional. 
Our convention is to suppress the dimension label on $I_n$
when we are dealing with a $D=4-2\eps$ integral.  We also suppress
the square brackets when the argument is unity.

\begin{figure}
  \begin{center}
    ~\epsfig{file=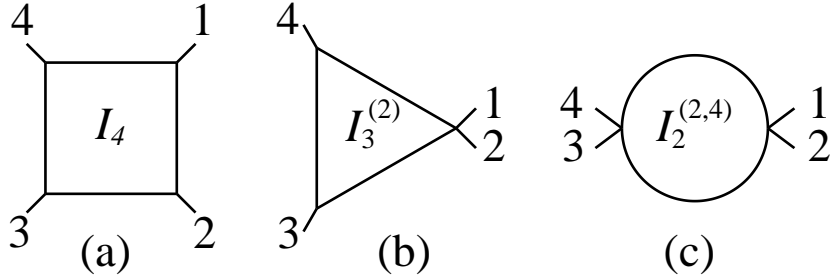,clip=}
    \caption[]{
      \label{FigLoopIntegral}
  Examples of labels on integral functions with four-point kinematics.
  The triangle $I_3^{(2)}$ in (b) is obtained by removing the internal
  propagator between external legs 1 and 2 in (a).  Similarly, the
  bubble $I_2^{(2,4)}$ in (c) is obtained by removing the internal propagator
  between legs 3 and 4 in (b).
}
  \end{center}
\end{figure}

Given a four-point amplitude, we have two conventions for labeling the
one-, two- and three-point integrals. The first, and more familiar one
is to explicitly give the kinematic invariant upon which the integral
depends. For example, $I_2(s)$ is the $(4-2\eps)$-dimensional bubble
that has an invariant mass square of $s=(k_1+k_2)^2$ flowing through
its external legs.  The second convention for labeling the lower point
integrals is to indicate, by a raised $(i)$ that the internal
propagator between legs $i-1$ (mod $4$) and $i$ has been removed.  If two
indices are present as in $(i,j)$ then two propagators have been
removed.  This follows the labeling convention of
refs.~\cite{IntegralRecursion} and is summarized in
\fig{FigLoopIntegral}.

For this paper it is sufficient to give explicit expressions for
the $(4-2\eps)$-dimensional one, two, three and four-point integrals
to $\Ord(\eps^0)$, but in some cases the all orders expressions are 
easy to obtain.

\subsection{The $D=4-2\eps$ scalar integral functions}

The one-point function is, in closed form,
$$
I_1 = m^{2-2\eps} { \Gamma(1+\eps) \over \eps (\eps - 1)} \, .
\equn\label{Tadpole}
$$
As expected from simple power counting, this is quadratically 
divergent and is singular for both $\eps = 0$ and $1$.

The bubble with an external kinematic invariant $s$ is 
$$
I_2(s) = I_2^{(2,4)} = I_2(0) 
+ 2 + x \log \left( {x-1 \over x+1} \right)
+ \Ord(\eps)\,,
\equn\label{EvalI2mass}
$$
where $x \equiv \sqrt{1 - 4m^2/s}$. 
For external massless kinematics, this may be written in closed form as as
$$
I_2(0) = m^{-2\eps} \, { \Gamma(1+\eps) \over \eps } \, .
\equn\label{BubbleFour}
$$

The scalar triangle integral is
$$
I_3(s) = 
I_3^{(2)} =
- {1\over 2s} \log^2 \left( {x+1 \over x-1} \right)
+ \Ord(\eps)\,,
\equn
$$
while the box with a uniform internal mass is 
$$
I_4(s,t) = -{1\over st} \left[
  H \Bigl( -{u m^2\over st},{m^2\over s} \Bigr)
  + H\Bigl(-{u m^2\over st},{m^2\over t}\Bigr)
\right] + \Ord(\eps) \,,  
\equn\label{BoxDef}
$$
where $u=-s-t$,
$$
\eqalign{
H(X,Y) &\equiv {2 \over x_+ - x_-} \biggl[
\ln \Bigl(1 - { X \over Y }\Bigr) \ln \Bigl({ -x_- \over x_+ }\Bigr)
\cr
&- {\rm Li}_2\Bigl({x_- \over y - x_+}\Bigr)
- {\rm Li}_2 \Bigl({x_- \over x_- - y}\Bigr)
+ {\rm Li}_2\Bigl({x_+ \over x_+ - y}\Bigr)
+ {\rm Li}_2\Bigl({x_+ \over y - x_-}\Bigr)
\biggr] \, , 
\cr
{\rm ~~~~with~~~~}
&
x_\pm = \half ( 1 \pm \sqrt{1-4X} )\, , 
\hskip 1 cm 
y = \half ( 1 + \sqrt{1-4Y} ) \,, 
}
\equn
$$
and the dilogarithm \cite{Lewin} is ${\rm Li}_2(x) = -\int_0^1 dt\;
\ln(1-xt)/t$.

For the case of a single massive external leg, with an invariant
mass-square $M^2$, we have
$$
I_4(s,t,M^2) = -{1\over st} \left[
     H \Bigl( -{u m^2\over st},{m^2\over s} \Bigr)
  +  H \Bigl(-{u m^2\over st},{m^2\over t}  \Bigr)
  -  H \Bigl(-{u m^2\over st},{m^2\over M^2}\Bigr)
\right] + \Ord(\eps) \,,
\equn\label{ExtMassBoxDef}
$$
where $u = M^2 - s - t$.  Through $\Ord(\eps^0)$, the box integrals
contained in $\K_4^{(i)}$ in \eqn{FiveGluonAmpl} may be obtained from
here and from \eqns{(\ref{HigherToLower}) and (\ref{DefineK})}; for example,
$I_{4}^{(1)} = I_4(s_{23},s_{34},s_{51})$.

The scalar pentagon integral appearing in five-point amplitudes
may be obtained from four-point integrals using 
the recursion formula \cite{VNV,IntegralRecursion}
$$
I_5 = {1\over 2} \sum_{i=1}^5 c_i I_{4}^{(i)}
+ \Ord(\eps) \, , 
\equn\label{PentagonInt}
$$
where 
$$
\eqalign{
  c_i = \sum_{j=1}^{5} S_{ij}^{-1} \, ,  \cr
}
\equn\label{CiDefinitions}
$$
and the $I_{4}^{(i)}$ are boxes with a uniform internal mass and a
single massive external leg.  The matrix $S_{ij}$ is
$$
 S_{ij} \equiv  m^2 - {1 \over 2}  p_{ij}^2\,,  \hskip .3 cm 
 {\rm with} \hskip .3 cm 
   p_{ii}\ \equiv\ 0\,, \hskip .3  cm {\rm and} \hskip .3 cm 
  p_{ij} = p_{ji} \equiv
  k_i+k_{i+1}+\cdots+k_{j-1}\quad{\rm for}\ i<j \, .
\equn\label{SDef}
$$
Similar formulas exist for arbitrary massive or massless 
kinematics \cite{IntegralRecursion}.

\subsection{The higher dimension integrals}
\label{HigherDimensionAppendix}

The higher dimension integrals are defined in \eqn{GeneralLoop} with
$D$ set to the appropriate value.  The relationship of these integrals
to the usual four-dimensional ones has been extensively discussed in
ref.~\cite{IntegralRecursion}.  For $n\le 6$ they can be written in
terms of the $(4-2\eps)$-dimensional integrals via the integral
recursion relations
$$
\eqalign{
  I_n^{D=6-2\eps} & = {1\over (n-5+2\eps) c_0} \biggl[
         2 I_n^{D=4-2\eps}
 - \sum_{i=1}^n c_i I_{n-1}^{(i),D=4-2\eps}
  \biggr] \, , \cr
  I_n^{D=8-2\eps} & = {1\over (n-7+2\eps) c_0} \biggl[
         2 I_n^{D=6-2\eps}
 - \sum_{i=1}^n c_i I_{n-1}^{(i),D=6-2\eps}
  \biggr]. \cr
}
\equn
\label{HigherToLower}
$$ 
where $c_i$ is defined in \eqn{CiDefinitions} (with the five-point kinematics
replaced by $n$-point kinematics)
and $c_0 = \sum_{i=1}^{n} c_i$.

Higher dimension integrals arise naturally when performing the cut
calculations described in this paper. 
In a typical cut calculation discussed in the text we obtain integrals
of the form
$$
\eqalign{
  I_n^{D=4-2\epsilon}[&f(p^\alpha, k_i^\alpha,\mu^2)] =\cr
 & i (-1)^{n+1} (4\pi)^{2-\epsilon}
  \int \frac{ d^{4}p }{(2\pi)^{4} }
  \frac{d^{-2\epsilon}\mu}{(2\pi)^{-2\epsilon}}
  \frac{f(p^\alpha, k_i^\alpha, \mu^2)}{(p^2 - \mu^2 - m^2) \ldots
          ((p-\sum_{i=1}^{n-1} k_i )^2 - \mu^2 - m^2) } \, ,
}
\equn\label{BetterLoop}
$$
where we have explicitly broken the $(4-2\eps)$-dimensional momentum
into a four-dimensional part, $p$, and a $(-2\eps)$-dimensional part, $\mu$.

Since we are using the Minkowski metric of negative signature and the
fractional vector has only space-like components, we define $\mu \cdot
\mu = - \mu^2$. Note that dimensional regularization can be thought of
as altering the mass appearing in the loop propagators.  This mass is
integrated over when we perform the fractional dimensional vector
integration: $\int d^{-2\epsilon}\mu$.

In general, odd powers of $\mu^\alpha$ cancel from one-loop integrals.
The $\mu^\alpha$ integration, of
\eqn{BetterLoop} is  formally in a sub-space that does not
overlap with any other of the momenta associated with the loop. For
this reason any contribution to the numerator that is odd in the
vector $\mu^\alpha$ will not contribute to the integral. Accordingly,
we shall only need to consider the cases where the integrand depends
on $\mu^\alpha$ only through $\mu^2$.
Note that we do not need to consider the more
general cases of $\mu^\alpha \mu^\beta$ etc., because in a full
amplitude all of the $(-2\eps)$-dimension Lorentz indices must be contracted
against each other.

Consider an integral of the form 
$$
\int {d^{4-2\eps} P \over (2\pi)^{4-2\eps}} \; (\mu^2)^r  f(p^\alpha, \mu^2)
 \, , 
\equn
$$
where $p$ and $\mu$ are the four- and $(-2\eps)$-dimensional components of $P$.
Breaking up the measure into the two types of components we have
$$
\eqalign{
 \int {d^4 p \over (2\pi)^4 }
  \int {d^{-2\epsilon}\mu \over (2\pi)^{-2\epsilon}}
                   \; (\mu^2)^r f(p^\alpha, \mu^2) 
  & = \int {d^4 p \over (2\pi)^4 } \int d\Omega_{-1-2\epsilon}
  \int_0^\infty  {d\mu^2 \over 2(2\pi)^{-2\eps}}
  \; (\mu^2)^{-1 - \eps + r}
  f(p^\alpha, \mu^2) \cr 
  & = {(2\pi)^{2r} \int d\Omega_{-1-2\epsilon} \over
    \int d\Omega_{2r-1-2\epsilon}}      
  \int {d^4 p \over (2\pi)^4 } 
  \int {d^{2r-2\epsilon}\mu \over (2\pi)^{2r-2\epsilon}} \; 
  f(p^\alpha, \mu^2) \cr 
  & = 
  - \eps (1-\eps) (2-\eps) \cdots (r-1-\eps)
  \; (4\pi)^{r} 
  \int {d^{4+2r- 2\eps} P \over (2\pi)^{4+2r-2\eps}} \; 
  f(p^\alpha, \mu^2) \,, \cr}
\equn
$$
where we used the independence of the integrand 
on the orientation of the vector $\mu^\alpha$ and 
the formula for solid angles in general dimensions
$$
\int d\Omega_n = \frac{ 2\,\pi^{\frac{n+1}{2}}  }
    {\Gamma \left(\frac{n+1}{2}\right)}\, .
\equn
$$
In particular, using the definition (\ref{GeneralLoop}), we find
$$
I_n^{D=4-2\epsilon}[\mu^2] = -\epsilon I_n^{D=6-2\epsilon} \, , 
\hskip 1 cm  \hbox{and} \hskip 1 cm 
I_n^{D=4-2\epsilon}[\mu^4] = -
\epsilon(1-\epsilon) I_n^{D=8-2\epsilon} \, .
\equn\label{DSEIntegrand}
$$
Note that although the loop momentum is shifted to higher 
dimension, the external momenta remain in four dimensions. 

An analogous method leads to the result,
$$
\int \frac{d^4 q}{(2\pi)^4}
\frac{d^{-2\eps}\mu}{(2\pi)^{-2\eps}}
q^{2r} f(q^2,\mu^2)
=
(-4\pi)^r \; (r+1)!
\int \frac{d^{4+2r-2\eps} Q}{(2\pi)^{4+2r-2\eps}}
 f(q^2,\mu^2) \,,
\equn
$$
which is useful when evaluating Feynman parameter integrals.

\subsection{Feynman parameter reduction}
\label{FeynmanParamAppendix}

As discussed in the text, tensor integrals may be 
evaluated using Feynman parameters. We make use of the following
formulae in the text.

After Feynman parameterization and the normal shift of momentum we
make use of the following,
$$
\eqalign{
\int {d^4 q\over (2\pi)^4}
 {d^{-2\epsilon}\mu \over (2\pi)^{-2\epsilon}}  \; 
 \frac{q^{\alpha_1} \cdots q^{\alpha_{2r+1}}}{(q^2 - \mu^2
- S_{ij} a_i a_j)^n} &=
  0 \, ,\cr}
\equn
$$
$$
\eqalign{
\int {d^4 q\over (2\pi)^4}
 {d^{-2\epsilon}\mu \over (2\pi)^{-2\epsilon}}  \;
\frac{q^\alpha q^\beta}{(q^2 - \mu^2 
- S_{ij} a_i a_j)^n}  &=
  {1\over 4} 
\int {d^4 q\over (2\pi)^4}
 {d^{-2\epsilon}\mu \over (2\pi)^{-2\epsilon}}  \; 
 \frac{q^2 \eta^{\alpha\beta}}{(q^2 - \mu^2 
- S_{ij} a_i a_j)^n}   \cr 
& = -{i\over 2}\, {(-1)^{n}  \over (4\pi)^{2-\eps}} \,\eta^{\alpha\beta}\,
 I_n^{D=6-2\eps} \, , \cr} 
\equn
$$
where $S_{ij}$ is the matrix defined in \eqn{SDef} with $n$-point kinematics.

After integrating out the loop momenta we obtain integrals of the form
$$
I_n^D[f(a_k)] \equiv  \Gamma(n-D/2)\int_0^1 d^na_k\
   \delta (1 - {\textstyle \sum_r} a_r)
     { f(a_k) \over \left[ \sum_{i,j=1}^n S_{ij} a_i a_j - i\varepsilon
            \right]^{n-D/2}}\, ,
\equn
$$
where $S_{ij}$ is defined in \eqn{CiDefinitions}. 
An integral reduction formula that is quite useful is
\cite{IntegralRecursion}
$$
  I_n^D[a_i]\ =\ {1\over2} \sum_{j=1}^n c_{ij}\ I_{n-1}^{(j), \, D}
   \ +\ {c_i\over c_0}\ I^D_n \, ,
\equn\label{LinearIntegralRelation}
$$
where
$$
  c_{ij}\ =\ S^{-1}_{ij} - {c_ic_j\over c_0}\, , 
\equn
$$
and $c_i$ is defined in \eqn{CiDefinitions} (with the five-point kinematics
replaced by $n$-point kinematics) and $c_0 = \sum_{i=1}^{n} c_i$.

\subsection{Common integral combinations}
\label{KandJ}

The following combination of integrals often appears in the amplitudes
of this paper.
$$
\eqalign{
  \J_n &\equiv I_n[m^2 + \mu^2] = I_n[m^2] + I_n[\mu^2] \cr
  &= m^2 I_n^{D=4-2\epsilon} [1] - \epsilon I_n^{D=6-2\epsilon} [1] \,, 
}
\equn\label{DefineJ}
$$
and
$$
\eqalign{
  \K_n &\equiv I_n[(m^2 + \mu^2)^2]
  = m^4 I_n[1] + 2m^2 I_n[\mu^2] + I_n[\mu^4] \cr
  &= m^4 I_n^{D=4-2\epsilon} [1]
  - 2m^2\epsilon I_n^{D=6-2\epsilon} [1]
  - \epsilon(1-\epsilon) I_n^{D=8-2\epsilon} [1] \,. 
}
\equn\label{DefineK}
$$

Using \eqn{HigherToLower} we can remove the explicit $\eps$ and $m^2$ 
dependence from these integral combinations. Specifically, we find
$$
\eqalign{
 & J_2(s) = {s\over 4} I_2(s) + {1\over 2} I_1 
  - {3\over 2}I_2^{D=6-2\eps}(s)\, ,  \cr
 & J_3(s) = {1\over 2} I_2(s) - I_3^{D=6-2\eps}(s) \, , \cr
 & J_4    = - {st\over 4u} I_4 - {s\over 2u} I_3(s) - {t\over 2u} I_3(t)
  - {1\over2} I_4^{D=6-2\eps} \,, \cr
  }
\equn\label{JNoEps}
$$
and
$$
\eqalign{
  K_4 & \equiv 
   \frac{3}{4} I_4^{D=8-2\eps}
  + \frac{s}{4u} I_3^{D=6-2\eps}(s)
  + \frac{t}{4u} I_3^{D=6-2\eps}(t)
  + \frac{s t}{8u} I_4^{D=6-2\eps}
  - \frac{s}{2u} J_3(s)
  - \frac{t}{2u} J_3(t)
  - \frac{s t}{4u} J_4\, .  \cr}
\equn\label{KNoEps}
$$
The forms in \eqns{(\ref{JNoEps}) and (\ref{KNoEps})} are more useful
for identifying the quadratic divergences.

\section{Fixing remaining ambiguities}
\label{FixingRemainingAppendix}

In this appendix we show that the two-step procedure, given in
\sec{FixAmbiguitySection}.2, uniquely fixes all cut-free functions in
the four-gluon amplitudes with massive loops.  Here, we fix the
coefficients of these functions using the known ultraviolet
divergences of amplitudes. In general, infrared mass singularities
also provide a means for fixing such coefficients.  

First we establish the set of cut free functions that may appear
in the amplitudes and then we explain why the two steps, in
\sec{FixAmbiguitySection}.2, fix the coefficients of all cut-free
functions.  For null momenta, $k^2 \rightarrow 0$, the following three
functions have no cuts in kinematic variables
$$
 I_1\, , \hskip 1truecm   I_2(k^2)\, , \hskip 1truecm
 \hbox{~~~and~~~}  I_2^{D=6-2\eps}(k^2) \, , 
\equn\label{CutFreeInts}
$$
where the integral functions $I_1$ and $I_2$ are scalar tadpole and
bubble functions in \eqns{(\ref{Tadpole}), (\ref{BubbleFour}) and
  (\ref{HigherToLower})}.  In the massless case ($m = 0$) these three
cut-free integrals identically vanish in dimensional regularization
and are irrelevant.  Observe also that, for $k^2 \ne 0$, the two-point
functions do have a cut in $k^2$, which means that only coefficients
of the tadpoles are cut-ambiguous. Thus, there are fewer cut-ambiguous
functions to consider for massive external legs.

First we show that there are no combinations of other integrals
which are cut free.  To do this we must establish the set of functions
that can appear in the amplitude.  Using a Passarino-Veltman (PV) type
reduction \cite{PV,Integrals}, one may re-express all tensor integrals
from the Feynman diagrams as linear combinations of $D=4-2\eps$ scalar
integrals.  Accordingly, we only need to consider scalar integral functions.

In general the coefficients of integral functions appearing in
amplitudes may depend on $\eps$.  An $\eps$ in a coefficient of a
divergent integral may lead to a rational function at $\Ord(\eps^0)$,
as in \eqn{PolyTerm}.  If the integral function combinations are
cut-free, this would complicate our analysis since one would need to
track the $\eps$-dependence.  For example, if a coefficient were
$\eps$ dependent one might obtain cut-free functions of the form
$$
C(\eps)\, I_2(0) = 
(a + b\eps + \cdots) \, m^{-2\eps} \, { \Gamma(1+\eps) \over \eps } 
= a \, m^{-2\eps} \, { \Gamma(1+\eps) \over \eps}  + b + \Ord(\eps) \, . 
\equn
$$
Whilst the coefficient $a$ could be fixed from knowledge of the
ultraviolet divergence (or from $m \rightarrow 0$ properties) of the
amplitude, without a detailed understanding of the coefficient
$C(\eps)$ we would not be able to fix the rational function, $b$. It
is thus convenient to eliminate all $\eps$-dependence from the
coefficients. To do so, we use an appropriate regularization scheme and
a slightly modified reduction algorithm.

In addition to altering the number of loop momentum dimensions,
conventional dimensional regularization also changes the number of
physical states; this introduces $\eps$-dependence at the level of the
Feynman diagrams.  Using the four-dimensional helicity scheme
\cite{Siegel,Long,KunsztFourPoint}, however, the number of physical
states is the same as in four-dimensions, so no $\eps$-dependence is
introduced in the diagrams.

We also use a systematic procedure for reducing tensor integrals which 
produces no $\eps$-dependence in the scalar integral coefficients%
\footnote{In the Passarino-Veltman paper \cite{PV}, this
  $\eps$-dependence is not explicit. Instead, it appears
  as constant terms not proportional to loop integrals. For
  example the $1/4$ in the reduction function, $C_{24}$, on p199 of
  this paper arises from an $\eps$ in a coefficient multiplying an 
  ultraviolet divergent integral.}.
This modified PV reduction is presented in the first appendix of
ref.~\cite{Factorization}; all $\eps$-dependence from the integral
reduction is absorbed into higher-dimension scalar integral
functions. Using the representation for $\J_n$ and $\K_n$
in \eqns{(\ref{JNoEps}) and (\ref{KNoEps})}, the four-point amplitudes
given in this paper may be explicitly expressed in this form.  (At any
point, the amplitudes may be re-expressed solely in terms of the more
conventional $D=4-2\eps$ integrals via the integral recursion formulas in
\eqn{HigherToLower}.)

We need to understand which linear combinations of integrals in a
gedanken calculation of an amplitude can produce a cut-free function.
For four-point amplitudes, this can be found from a direct inspection
of the integral functions in \app{IntegralsAppendix}.  (This analysis
is similar to the one performed for massless amplitudes in
ref.~\cite{SusyOne}.)

Putting aside the higher-dimensional integrals, we first consider the
$D=4-2\eps$ integrals in the first part of \app{IntegralsAppendix}.  The
coefficients of those integrals with kinematic dependence are fixed by
the cuts because each of the functions contains a logarithm or
dilogarithm unique to it: the bubbles contain single logarithms, the
triangles squares of logarithms, and the box dilogarithms that tangle
the $s$ and $t$ channels.  Consequently no linear combination of these
integrals can be formed which is cut free.

To show that the higher dimension integrals with nontrivial kinematics
can be distinguished with the all orders in $\eps$ cuts, we make use of
the integral recursion relation \eqn{HigherToLower}. This relationship
between higher dimension integrals and $D=4-2\eps$ integrals
introduces a distinct $\eps$-dependence to coefficients.  For example,
by eliminating $I_4^{D=6-2\eps}$ in favor of $D=4-2\eps$ integrals
with \eqn{HigherToLower}, a factor of $(-1+2\eps)^{-1}$ is introduced
into the coefficient of the $D=4-2\eps$ box.  In reducing a
$D=8-2\eps$ dimensional box to $D=4-2\eps$ dimensional ones, an
additional factor of $(-3+2\eps)^{-1}$ is obtained.  Since the cuts
are computed to all orders in $\eps$, the cuts detect the full
analytic structure of these coefficients.  Similarly, to distinguish
between $I_3^{D=6-2\eps}$, $I_2^{D=6-2\eps}$ and $I_2^{D=4-2\eps}$, we
note that after applying the integral recursion formula
(\ref{HigherToLower}), we obtain $D=4-2\eps$ scalar bubbles,
proportional to $(-2 +2\eps)^{-1}$, $(-3+2\eps)^{-1}$ and $1$,
respectively.  Given an amplitude expressed as a linear combination of
$D=4-2\eps$ and higher dimension integrals, with coefficients having no
$\eps$-dependence, the coefficients of the higher dimension integrals
with nontrivial kinematics
are therefore uniquely specified by the cuts.

Thus, the only functions with no cuts which may appear in an amplitude
are the ones listed in \eqn{CutFreeInts}, so 
any four-point gauge theory amplitude will be of the form%
$$
A_4 = \hbox{Cut-constructed part} + 
\lim_{k^2 \rightarrow 0} \Bigl( d_1\, I_1 + d_2\, I^{D=6-2\eps}_2(k^2) 
+ d_3 \, I_2(k^2) \Bigr) \, . 
\equn
$$
The coefficients $d_i$ may contain a $1/k^2$ so it is convenient to to
write the cut ambiguous parts as a limit.  A single power of $1/k^2$
is the worst that one encounters in a one-loop gauge theory calculation.  (For
a mass in the loop the $k^2 \rightarrow 0 $ limit is smooth and no
infrared divergences develop.)

Expanding the integral $I_2^{D=6-2\eps}[1]$, defined in \eqn{GeneralLoop},
with respect to the external momentum we have
$$
\eqalign{
I_2^{D=6-2\eps}(k^2) & =  m^{2-2\eps} { \Gamma(1+\eps) \over \eps (\eps-1) }
+ {1\over 6} k^2  m^{-2\eps} { \Gamma(1+\eps) \over \eps } + \Ord(k^4) \cr
&  = I_1 
+ {1\over 6} k^2  I_2(0) + \Ord(k^4) \, , \cr}
\equn
$$
so $I_2^{D=6-2\eps}(0)$ contributes only as a linear combinations of 
$I_1$ and $I_2(0)$.
Thus, there are only two independent cut-free integral functions which
may enter an amplitude with a uniform mass in the loop. Hence, the
amplitude must have the form
$$
A_4 = \hbox{Cut-constructed part} + 
\tilde{d}_1 \, I_1 + \tilde{d}_2 \, I_2(0) \, .
\equn\label{LateAmbiguity}
$$

Since there are only two cut-free functions, $I_1$ and $I_2(0)$, 
and two types of ultraviolet divergences, quadratic and logarithmic, 
we may fix the coefficients $\tilde{d}_1$ and $\tilde{d}_2$ by making 
the amplitude have 
the correct total ultraviolet divergence.  This explains the two step 
procedure of \sec{FixAmbiguitySection}.

Following the discussion for the massless case in ref.~\cite{SusyOne}
one may extend the above arguments to an arbitrary number of legs.  (One
difference is that when working to all orders in $\eps$, scalar pentagon
integrals can no longer be eliminated from the set of integrals in
terms of which the answer is expressed.)

\section{Rules for $\mu^\alpha$ and $\s\mu$}
\label{MuSlashEtc}

The identification of dimensional regularization with the introduction
of an integrated mass-like vector, $\mu^\alpha$, necessitates some
discussion. 
To begin, we have a $(4-2\epsilon)$-dimensional vector,
$Q^\alpha$, which we can express as a sum of a four
dimensional piece $q^\alpha$ and a remainder $\mu^\alpha$,
$Q^\alpha = q^\alpha + \mu^\alpha$. It follows that $\s{Q} =
\s{q}+\s\mu$. Here we discuss the properties of
$\s\mu$.

Firstly, some definitions. We abide by the usual conventions for the
Dirac algebra,
$$
 \left\{
    \gamma^\alpha , \gamma^\beta
  \right\} = 2 \eta^{\alpha\beta} \, , \hskip 2 cm 
  \gamma^{\alpha\dagger} \gamma^0 = \gamma^0 \gamma^{\alpha}.
\equn
$$
In this equation, $\alpha$ and $\beta$ are $(4-2\epsilon$)-dimensional Lorentz
indices and the metric is $\eta^{\alpha\beta} =
{\rm diag} (+,-,-,-,-,\ldots)$. It follows that,
$$
  2 Q\cdot Q = \left\{
    \s{Q} , \s{Q}
  \right\} =
      \left\{ \s{q} , \s{q} \right\}
  +   \left\{ \s{\mu} , \s{\mu} \right\}
  + 2 \left\{ \s{q} , \s{\mu} \right\} 
  = 2 q\cdot q + 2 \mu\cdot \mu + 4 q \cdot \mu
  = 2 ( q^2 - \mu^2 ) \, .
\equn
$$
The cross-term vanishes because we have formally chosen $\mu^\alpha$
to be in a sub-space orthogonal%
\footnote{This choice does not correspond to the one made in the
dimensional reduction \cite{Siegel} regularization scheme.}
to the four-dimensional subspace containing $q^\alpha$.    Without
this choice, terms of the form $\mu \cdot k$ would exist, significantly 
complicating our calculations.
The minus sign in going from $\mu\cdot \mu$ to $\mu^2$ is from the
metric. For the familiar four-dimensional vector, as is conventional,
we write $q\cdot q \equiv q^2$.

When  the $(4-2\epsilon)$-dimensional vector is
null,
we find that the four-dimensional vector
effectively becomes `massive':
$$
  Q\cdot Q = 0 \Rightarrow q^2 = \mu^2 \,, 
\equn
$$
where we denote a
$(4-2\eps)$-dimensional momentum by a capital letter and a
four-dimensional momentum by a lower-case letter.  The quantity
$\mu^2$ is always integrated over.

Since the metric is diagonal, $\s\mu$ freely {\em anti-commutes} with
four-dimensional $\gamma^\alpha$ matrices,
$$
  \left\{ \s{q} , \s\mu \right\} = 0\,.
\equn
$$
This has an interesting consequence with respect to manipulations with
$\gamma_5$.  We use the conventions of 't~Hooft and Veltman
\cite{DimensionalRegularization,CollinsBook}, and adopt the arbitrary
dimension definition, $ \gamma_5 = i \gamma^0 \gamma^1 \gamma^2
\gamma^3$.  As usual, $\gamma_5$ {\em anti-commutes} with all the
four-dimensional Dirac matrices:
$$
  \left\{ \s{q} , \gamma_5 \right\} = 0 \, .
\equn
$$
With respect to $\s\mu$, however, $\gamma_5$ freely {\em commutes},
$$
  \left[ \s{\mu} , \gamma_5 \right] = 0\, ,
\equn
$$
since $\gamma_5$ is defined in this prescription as a product of an
even number of four-dimensional Dirac matrices.  The
$(-2\eps)$-dimensional component, $\s{\mu}$, commutes freely with the
helicity projection operator, $\omega_\pm \equiv \frac{1}{2} ( 1 \pm
\gamma_5 ),$
$$
  \omega_\pm \,\s{q} = \s{q} \,\omega_\mp \, , 
  \mbox{~~~~but,~~~~}
  \omega_\pm \, \s{\mu} = \s{\mu} \, \omega_\pm \, .
\equn\label{HelicityProjection}
$$

In our dimensionally regularized cut calculations the momentum of cut
fermion lines is to be considered $(4-2\epsilon)$ dimensional.  We
will borrow the bra and ket symbols to represent the spinors, but as
helicity is not a good quantum number we
shall not label them with a helicity.  (Since we always
sum over all states across the cut, there is anyway no need to define
a $(4-2\epsilon)$ helicity notion.)  These spinors obey the
conventional Dirac equation,
$$
  \bra{Q} \s{Q} = \bra{Q} m \, , 
 \hskip 2 cm 
  \bra{Q} \s{q} = \bra{Q} ( m - \s{\mu} ) \,. 
\equn
$$
To handle trailing fermions we will use the same expression but in
such a case the momenta will be explicitly negative, $\s{q} \ket{-Q} =
-( m + \s{\mu} ) \ket{-Q}$. We adhere to the
convention that the argument momentum is the momentum flowing in the
direction of the fermion arrow.

In sewing the $(4-2\epsilon)$-spinors together (across the
cut) we implicitly sum over the two spin degrees of freedom, that is
$$
  \ket{Q} \bra{Q} = \s{Q} + m = \s{q} +\s{\mu} + m \, , 
  \mbox{~~~and~~~}
  \ket{-Q} \bra{-Q} = -\s{Q} + m = -\s{q} -\s{\mu} + m \, .
\equn\label{SpinSum}
$$
As mentioned in the text, this notation glosses over the
distinction of spinors and anti-spinors and can introduce an overall
minus sign that needs to be put back by hand.

Having kept a four-dimensional definition for $\gamma_5$, we can apply
spinor-helicity methods \cite{SpinorHelicity} to compress tree
amplitudes.  However, some care is required in applying the identity
(\ref{PolSlash}).  Whereas this is the conventional rule for
substituting for a slashed polarization vector, it is only a valid
identity when the Dirac algebra on at least one side is
four-dimensional. For example, for $\bra{Q} \s{\pol}^\pm \s\mu
\s{\pol}^\pm \s\mu \ldots$, before applying the rule (\ref{PolSlash}) 
we must first
anti-commute each of the $\s\mu$ to one end of the spinor line.  In
this way we pick-up factors of $-\s\mu\s\mu = \mu^2$, just as we would
combine conventional mass contributions from the fermion propagator.
Once we have two neighboring polarization vectors, $\s{\pol}^\pm(k;r)
\s{\pol}^\pm(p;q)$, we are free to apply the rule (\ref{PolSlash}).

\end{document}